\documentclass[journal=jacsat,manuscript=article]{achemso}

\usepackage[version=3]{mhchem} % Formula subscripts using \ce{}
\usepackage{algorithm}
\usepackage{algpseudocode}
\usepackage{hyperref}
\usepackage{xcolor}
\usepackage{multirow}
\author{Jie Li}
\affiliation{ Pitzer Center for Theoretical Chemistry, Department of Chemistry}

\author{Xingyi Guan}
\affiliation{ Pitzer Center for Theoretical Chemistry, Department of Chemistry}

\author{Oufan Zhang}
\affiliation{ Pitzer Center for Theoretical Chemistry, Department of Chemistry}

\author{Kunyang Sun}
\affiliation{ Pitzer Center for Theoretical Chemistry, Department of Chemistry}

\author{Yingze Wang}
\affiliation{ Pitzer Center for Theoretical Chemistry, Department of Chemistry}

\author{Dorian Bagni}
\affiliation{ Pitzer Center for Theoretical Chemistry, Department of Chemistry}

\author{Teresa Head-Gordon}
\affiliation{ Pitzer Center for Theoretical Chemistry, Department of Chemistry}
\alsoaffiliation{Departments of Bioengineering and Chemical and Biomolecular Engineering, \\ University of California, Berkeley, CA, USA}
\email{thg@berkeley.edu}

\title
  {Leak Proof PDBBind: A Reorganized Dataset of Protein-Ligand Complexes for More Generalizable Binding Affinity Prediction}

\abbreviations{IR,NMR,UV}

\begin{document}


\begin{abstract}
\noindent
The majority of machine learning scoring functions used in drug discovery for predicting protein-ligand binding poses and affinities have been trained on the PDBBind dataset. However, it is unclear whether these new scoring functions are actually an improvement over traditional models since often the training and test sets are cross-contaminated with proteins and ligands with high similarity, and hence they may not perform comparably well in binding prediction of unrelated protein-ligand complexes. In this work we have carefully prepared a new split of the PDBBind data set to control for data leakage, defined as proteins and ligands  with high sequence and structural similarity. The resulting leak-proof (LP)-PDBBind data is used to retrain four popular SFs: AutoDock Vina, Random Forest (RF)-Score, InteractionGraphNet (IGN), and DeepDTA, to better test their capabilities when applied to new protein-ligand complexes. In particular we have formulated a new independent data set, BDB2020+, by matching high quality binding free energies from BindingDB with co-crystalized ligand-protein complexes from the PDB that have been deposited since 2020. Based on all the benchmark results, the retrained models using LP-PDBBind consistently perform better, with IGN especially being recommended for scoring and ranking applications for new protein-ligand systems. 
\end{abstract}

\section{Introduction}
Many physics-based and machine-learned scoring functions (SFs) used to predict protein-ligand binding energies have been trained on the PDBBind dataset.\cite{pdbbind} The PDBBind dataset is a curated set of $\sim$20K  protein-ligand complex structures and their experimentally measured binding affinities, in which the "general" and "refined" data subsets are typically used for training and/or validation, and a separate "core" set is used for testing, in the development of new SFs. Significantly, however, the majority of the core data records in the PDBBind dataset have identical proteins and/or ligands with that found in the general and/or refined sets. As such, many empirical SF models have been trained with significant data leakage, and thus their reported performance on the core set is only a true measure of new protein-ligand complexes with high similarity, but will inevitably have limited transferability to low-similarity ligand-proteins scenarios.

The definition of data leakage is not as literal a scenario in which the \textit{exact} occurrence of input representation occurs in both training and test sets. Instead data leakage here refers to highly similar protein or ligands measured by sequence and chemical similarity and similar structure occurring in both the training and testing datasets. This will lead to artificially high evaluation results that do not fully reflect the performance of the models in real world applications for more diverse protein-ligand complexes.\cite{generalization} Therefore, the most standard data splitting protocol in the machine learning community, which is to randomly split the dataset into train, validation and test subsets will inevitably introduce significant data leakage issue. Given that many SFs training on PDBBind do not carefully investigate the data similarity issue, whether these new SFs have truly surpassed traditional SFs in actual predictive performance is still an open question, or how well any given SF performs on new protein-ligand binding applications.\cite{accuracy_or_novelty,similarity1,similarity2,deeplearning_vs_traditional_docking} It is already known that any SF model can be overtrained so that it has exceptional performance on the training dataset, a problem which can be mitigated by several known regularization strategies to provide better generalization to an independent test set.\cite{Haghighatlari2020,LeePing} Equally important, however, is to control for data leakage into the test set itself, without which can lead to false confidence in predictive capacity when the training dataset has high similarity to the test set, but manifests as poor generalization when the sequence or structure similarity is low.\cite{similarity1} 

Two typical solutions have been taken to mitigate this data leakage issue: one is to perform data splits based on scaffolds, protein families or protein clusters, and the other is to perform time-based cutoff which was proposed on earlier versions of PDBBind\cite{RFScorev3}, and was suggested by the authors of the EquiBind model on the PDBBind 2020 release\cite{equibind}. The scaffold or target based splitting can make sure the proteins in the complex are not similar between training and testing datasets, but they usually do not take the ligand similarities into account. The idea of a time-based splitting mimics a "blind test" setting, by which the model can only be trained with data released in year 2019 or earlier, and predictions are made with data after year 2019, so that the test data will never be covered by training data. However, since new drugs are being developed that can interact with popular protein targets that have been established for years,\cite{new_drug_old_target, new_drug_old_target2} and existing drug molecules may also be tested on new proteins\cite{drug_repurposing}, there are still frequent encounters with almost identical proteins or ligands in the latest experiments with earlier assays. As such, a time based splitting of the dataset is still not an ideal solution. We note that other data-splitting methods\cite{Datasail2025,Graber2025} have been reported in 2025 without attribution to our work on the Arxiv posted in 2023.\cite{li2023LP}

For the purpose of retraining SFs that can either truly generalize, or just as importantly to anticipate when it will not work well when similarity of protein-ligand complexes is low, we build a new data split to mitigate data leakage of highly similar sequence and structures of both proteins and ligands in the training and test set. In this work we reorganize the PDBBind data into new train, validation and test datasets which we call Leak Proof PDBBind (LP-PDBBind), by minimizing sequence and chemical similarity of the proteins and ligands between the datasets as well as making sure the protein-ligand structural interaction patterns are different among the data splits as well. Our new data splitting can be regarded as an improvement over just a protein scaffold split by providing similarity control on the ligands as well, which we hypothesize will decrease the chance of a model making predictions by memorizing ligand structures.\cite{gorantla2023proteins,volkov2022frustration} 

We use the new split for the development of new versions of several SFs, including AutoDock Vina\cite{autodock_vina}, RF-Score\cite{rfscore}, IGN\cite{IGN}, and DeepDTA\cite{deepDTA}. We note that DataSail\cite{Datasail2025} and CleanSplit\cite{Graber2025}, both reported in 2025, also retrained ML models but without attribution to our 2023 Arxiv paper.\cite{li2023LP} As we showed before, this provides a standardized way to train and benchmark different SFs with the exact same train, validation and test data. Furthermore, in order to provide a true independent benchmark for the new SFs that result from retraining using the new LP-PDBBind split, we created a new evaluation dataset, BDB2020+. BDB2020+ is compiled based on entries into the BindingDB dataset\cite{bindingdb1} that were deposited after 2020, and further filtered according to the same similarity control criteria used for the development of the new LP-PDBBind. We note that DataSail\cite{Datasail2025} also used our BDB2020+ data, which we generously provided to the community in the THG Lab github, but they used it without attribution.\cite{li2023LP} As a further test of ranking power, we additionally prepared two sets of experimental binding affinity data for different ligand complexes of the SARS-CoV-2 main protease (Mpro)\cite{mpro_inhibitors} and epidermal grow factor receptor (EGFR)\cite{EGFR}, neither of which was included in the LP-PDBBind training  dataset (although Mpro has similar SARS-CoV-1 protease proteins in PDBBind). Finally, we perform two control evaluated the performance of retrained SFs on docked structures using the old Vina SF to mimic a more typical drug discovery scenario.

The results show that all SFs trained on PDBBind show better performance on the test set due to data leakage, while the retrained models using the LP-PDBBind split are found to give a more true measure of performance. Furthermore, the newly trained SFs in general now perform much better on the completely independent BDB2020+ data set, showing that the new split can help with better generalization. Furthermore, even though the models were trained using co-crystal structures, improvements are also seen on redocked structures, thus demonstrating the applicability for the retrained SFs in actual practice. Hence, we believe the LP-PDBBind split provides a better way of utilizing the existing PDBBind dataset, and provides a more realistic and meaningful benchmark to help develop higher quality and more generalizable SFs in the future.

\section{Methods}

\subsection{Data cleaning of PDBBind v2020}
We have cleaned the PDBBind v2020 data to remove covalent ligand-protein complexes, the low populations of drug molecules with underrepresented chemical elements, and complexes with steric clashes; this is defined as clean level 1 (CL1). In addition to CL1, the CL2 level of cleaning aimed for consistent measures of binding free energies by either converting $K_d$ or eliminating data reported as $\mathrm{IC_{50}}$. All of the results reported here were based on training on CL1 data and testing on CL2. A more recent and thorough cleaning procedure, the HiQ-Bind workflow, has also been applied to PDBBind v2020 in that study.\cite{Wang2025} 

It is important to treat covalent and non-covalent binders separately, because most existing algorithms that predict protein-ligand binding primarily focus on non-covalent interactions. As far as we are aware, there has been no systematic study of whether the binders in PDBBind are covalent or non-covalent. Relying on the CovBinderInPDB \cite{guo2022covbinderinpdb} repository, we have identified covalent binders in the PDBBind dataset. The ligand names were extracted from the PDB files by comparing the minimum distance between any atom from the ligands in PDBBind database and residues in the PDB files downloaded from the RCSB database. If the minimum distance is less than 1Å, the matched residue name was compared with the record in CovBinderInPDB to identify whether the ligand is a covalent binder. If the minimum distance is more than 1Å or the residue name did not match the record in CovBinderInPDB, the structures were manually checked to identify whether the ligands are covalent binders or not. Ultimately 893 covalent binders and 18550 non-covalent binders were identified in the PDBBind dataset (see Supplementary Information). We also eliminated protein-ligand complexes with steric clashes, defined as a heavy-atom distance between protein and ligand that is below 1.75 Å (see Supplementary Information). In total there are 11,513 protein-ligand complexes in the training set, 2,422 in the validation set, and 4,860 in the test set before merging to create the new split (see Supplementary Table S1).

\subsection{New splitting of PDBBind dataset}
We have formulated a new splitting of the PDBBind dataset to minimize the similarity between training, validation and test data as much as possible in order to eliminate the risk of data leakage. Ligand similarities were calculated as the Dice similarity\cite{dice_sim} between a pair of ligands based on Morgan fingerprints of the ligands using 1024 bits\cite{ECFP}. Similarities for proteins were calculated as the percentage of matched number of residues over the length of aligned sequences after sequences were aligned using the Needleman-Wunsch alignment\cite{needleman_wunsch}. More details on the calculation of similarity for proteins and ligands and data splitting procedure are given in the Supplementary Information. The total amount of protein-ligand data in the training, validation and test data after the new splitting are 11513, 2422 and 4860, respectively, and the data for each category is reported in Supplementary Table S1. The new splitting procedure ensures the data in the training set has a maximum protein sequence similarity of 0.5 and maximum ligand similarity of 0.99 to any data in the validation or test datasets, and the maximum similarity between any validation and test data is 0.9 for protein sequence similarity and 0.99 for ligand similarity. The similarity cutoffs are carefully chosen to reach a balance between the dis-similarity of train and test datasets and the amount of data in each subset. A higher similarity cutoff for the ligands was used compared to the protein similarity cutoff because similar ligands might interact differently with different types of proteins, which is useful for a ML model to learn.

Additionally, we explored the protein-ligand interaction similarities using proteo-chemometrics interaction fingerprints\cite{fingerprint} and validated the exclusion of highly similar interaction patterns in the dataset using our splitting procedure. The proteo-chemometric interaction fingerprints extend the connectivity fingerprints (ECFPs) by the interactions between ligand atoms and nearby residues from the protein in
3D space, and maps the interaction patterns into a fixed size integer vector with a length of 256. The pairwise interaction fingerprint similarities were then calculated using a weighted Jaccard similarity score $$S_{\mathrm{IFP}}(X, Y)=\frac{\sum_i(\min(X_i, Y_i))}{\sum_i(\max(X_i,Y_i))}$$
where $X$ and $Y$ are the interaction fingerprints for two complexes. As we will show later, our splitting procedure also ensures distinct separation of data in the train, validation and test subset of PDBBind in terms of interaction fingerprints.

\subsection{Retrained Models using the new split}
The primary deliverable from this study are new SFs for protein-ligand binding prediction. We have considered both classical SFs as well as machine learned SFs. The commonly used AutoDock Vina\cite{autodock_vina} is designed to model intermolecular interactions or missing free energy components, but could benefit from better training data for parameterization of their semi-empirical or empirical functions.\cite{autodock4,autodock_vina,glide1,glide2,xscore,OpenFF1,OpenFF2}. Data-driven MLSFs by contrast are less reliant on physical interaction modeling and are far more dependent on experimental information\cite{knowledge_SF,statistical_potentials,PMF}, culminating in the current development and use of sophisticated machine learning (ML) models\cite{rfscore,IGN,pignet,rtmscore,deepDTA,holoprot,tankbind,GIGN} whose much larger parameter spaces are optimized on large, high quality datasets. Some representative MLSF models we consider include Random Forest Score (RF-Score)\cite{rfscore} that is based on atom-pair distance counts, InteractionGraphNet (IGN)\cite{IGN} that uses graph neural network (GNN) to represent raw 3-dimensional protein and ligand structures, and DeepDTA\cite{deepDTA} that predicts binding affinities using amino acid sequences of the proteins and SMILES strings of the ligands. More recent models have been proposed that achieve better accuracy at binding affinity prediction, including RFScore-V3 \cite{RFScorev3} that incorporates terms from AutoDock Vina, and also PIGNet\cite{pignet}, RTMScore\cite{rtmscore} and GIGN\cite{GIGN} which are more advanced GNN-based models.

\subsection{Compilation of the BDB2020+ dataset}
Because many of the recent SFs have been trained on PDBBind, which means utilizing a subset of PDBBind as the test dataset risks data leakage, we created a new benchmark dataset that is independent of PDBBind. To fulfill the need for a fair benchmark, we looked for the records deposited in BindingDB\cite{bindingdb1,bindingdb2} after year 2020 until 2023. Given that many data in BindingDB overlap with the PDBBind dataset which contains data up to year 2020, we have only looked at data in BindingDB deposited after year 2020 to minimize the chance of data leakage. BindingDB is one of the largest public binding affinity repositories, which also provides additional experimental conditions including assay information, pH and temperature. However, it does not guarantee each record has an associated 3D complex structure. 

We have developed a workflow to match complex structures in RCSB PDB with records in BindingDB. Since a complex may contain multiple ligands, but only one matches the record in BindingDB, we selected the ligand in the PDB that has the best structural match with the SMILEs provided in BindingDB, and ensuring that the number of heavy atoms is exactly the same. We then used rdkit\cite{rdkit} to reassign bond orders to the extracted ligands using the BindingDB SMILES as reference. This step was necessary because bond orders are usually not present in a PDB file and are typically inferred from local atomic geometries, which sometimes result in unreasonable bonding structures; thus the bond order reassignment step ensures the rationality of the processed structures.

Additionally, any chain that is within 5 Å of the ligand was compared with the interacting chain sequence in BindingDB record. A reliable match was only made when the consecutive aligned residues are exactly the same. The reason to keep strict alignment criterion is that if the protein contains mutations, the binding affinity might change significantly, in which case the BindingDB record will not represent the true binding affinity for the complex structure in the PDB and is not usable for the benchmark. After discarding all unmatched data, we obtained 130 data records, out of which 115 contains accurate binding affinity data, and defines the BDB2020+ test dataset. The flowchart of the building process of BDB2020+ dataset is illustrated in Supplementary Figure S1.

We have further prepared the redocked structures using the original AutoDock Vina for each protein-ligand complex in the BDB2020+ dataset to evaluate the performance of the retrained SFs on docked poses instead of co-crystal structures obtained from experiments, because the co-crystal structures are usually unavailable in real-world applications. Specifically, for each data in BDB2020+ dataset, we run AutoDock Vina docking to generate 30 poses with exhausiveness=64. The pose with the lowest Vina score and the pose that is most similar to the cocrystal structure in terms of symmetry-corrected RMSD \cite{CanonizedRMSD} were both retained. However, any data with the lowest RMSD to the co-crystal structure larger than 2 Å were excluded to prevent the influence of bad docking, resulting a total of 104 data remained for the redocked set of BDB2020+ that is closest to the co-crystal structure.

\subsection{Preparation of two target-specific datasets}
Finally, we have prepared two additional datasets of protein-ligand complexes with the same protein and different ligands so that we could evaluate the ranking accuracies of the SFs before and after retraining using LP-PDBBind. The first dataset is based on the SARS-CoV-2 main protease (Mpro) for which a wide variety of potential Mpro inhibitors have been developed.\cite{mpro_inhibitors}  We have manually extracted published co-crystal structures of Mpro with a number of non-covalent inhibitors\cite{mpro1,mpro2,mpro3,mpro4,mpro5,mpro6,mpro7,mpro8,mpro9,mpro10,mpro11,mpro12,mpro13}, and prepared a dataset containing 40 structures and corresponding experimental binding affinity measurements. The second dataset involves the epidermal growth factor receptor (EGFR), which is a receptor tyrosine kinase related to multiple cancers including lung cancer, pancreatic cancer and breast cancer.\cite{EGFR} Similarly, we have selected 23 representative non-covalent protein-ligand complex structures of EGFR with binding affinities taken from BindingDB.\cite{egfr1,egfr2,egfr3,egfr4,egfr5,egfr6,egfr7,egfr8,egfr9,egfr10,egfr11,egfr12,egfr13,egfr14,egfr15} The two dataset were selected as two representative
scenarios of using the SFs on a similar or different protein-ligand complex than what was
included in the training dataset.

\subsection{Retraining Scoring Functions with LP-PDBBind}
The LP-PDBBind data for non-covalent binders of PDBBind were used to retrain AutoDock Vina\cite{autodock_vina}, IGN\cite{IGN}, the 2010 RF-Score\cite{rfscore} and DeepDTA\cite{deepDTA}. These models cover a wide range of different approaches for scoring, and are representative of different dimensions of the SF space of models. AutoDock vina is a CSF that contains molecular interaction terms consisting of a van der Waals-like potential (defined by a combination of a repulsion term and two attractive Gaussians), a nondirectional hydrogen-bond term, a hydrophobic term, and a conformational entropy penalty, all of which are weighted by empirical parameters. The other three methods belong to MLSF category but vastly differ in their feature set. The RF-Score model is based on random forest regression that predicts binding affinities from the number of occurrences of a particular protein-ligand atom type pair interacting within a certain distance range using 3D structures.  InteractionGraphNet (IGN) utilizes a graph neural network operating on the 3D complex structures, and the node and edge features are straightforward information about atoms and bonds, including atom types, atom hybridization, bond order, etc. Finally, DeepDTA is a Y-shaped 1-dimensional convolutional neural network that takes in protein sequences and ligand isomeric SMILES strings as input and outputs the predicted binding affinities, making it a model that does not rely on the exact 3D structure of the complex.
 
Details on the retraining of these models are described in the Supplementary Information. In short, AutoDock vina was retrained by optimizing its six empirical parameters by minimizing the mean absolute error between predicted and experimental binding affinities using the Nelder-Mead optimization algorithm\cite{nedler_mead_opt}. Both IGN and DeepDTA were retrained by minimizing the mean squared error using gradient descent method as implemented in PyTorch\cite{pytorch}. For RF-Score, the optimization was driven by variance reduction between prediction and experimental affinities in each decision tree using the scikit-learn implementation\cite{scikit-learn}.  After retraining, the new AutoDock Vina, RF-Score, IGN, and DeepDTA model performances are compared with the old models as tested on the non-covalent LP-PDBBind test set using the CL2 data, the BDB2020+ new benchmark data, the Mpro and EFGR applications, and finally evaluated on the redocked structures in BDB2020+ dataset.

% Because MLSF are especially sensitive to well-curated data, we undertook some additional cleaning steps of the PDBBind data set. In the splitting process described above, ligand similarity might still exceed 0.99 for data from different categories. Therefore, any data in the combined training set that has ligand similarity greater than 0.99 to any other data in the validation or test set were discarded altogether. The resulting number of data in the training, validation, and test set were 11513, 2422 and 4860, respectively, after this cleaning step. 

\section{Results}
\subsection{Analysis of PDBBind Splittings}
\begin{figure}[H]
\begin{center}
\includegraphics[width=0.8\textwidth]{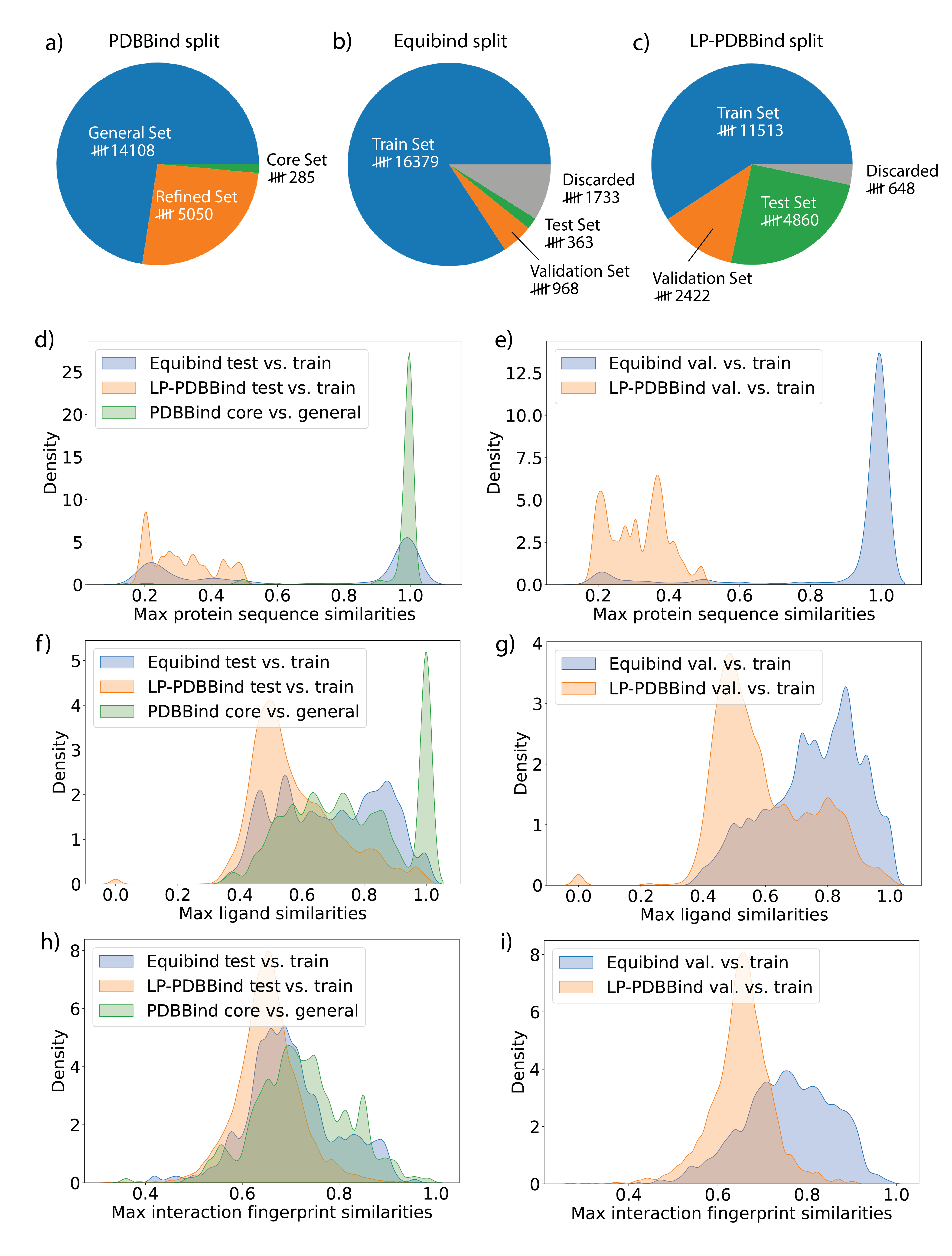}
\vspace{-7mm}
\end{center}
\caption{\textit{Data statistics under different splits of the PDBBind dataset.} Number of data in different splits of the PDBBind dataset: (a) general, refined and core set in original PDBBind split;  (b) train, validation, test set and discarded data in Equibind split; (c) train, validation, test set and discarded data in LP-PDBBind. Comparison of maximum protein sequence similarities between test set and train set (or core set and general set in original PDBBind split) (d) or between validation set and train set under Equibind split and LP-PDBBind(e). Comparison of maximum ligand similarities between test set and train set (or core set and general set in original PDBBind split) (f) or between validation set and train set under Equibind split and LP-PDBBind(g). Comparison of maximum interaction fingerprint similarities between test set and train set (or core set and general set in original PDBBind split) (h) or between validation set and train set under Equibind split and LP-PDBBind(i). }

\label{fig:pdbbind-data-stats}
\end{figure}

Data distributions of PDBBind under the original split (general set/refined set/core set), Equibind Split (train/validation/test) and LP-PDBBind (train/validation/test) are provided in Figure \ref{fig:pdbbind-data-stats} a-c. LP-PDBBind has significantly extended the size of the test set compared with the original PDBBind split and the more recent Equibind split. A bigger test set provides more accurate evaluation of model performance when the model is applied to data that has not been trained on.  By contrast, the number of data in the training set is much smaller in LP-PDBBind. As we will show later, the shrinkage in the training set is necessary, because it keeps the similarity with validation/test set sufficiently low. Compared with the Equibind split, we have also decreased the number of discarded data, because we only discard data when they are highly similar to any other data in train, validation or test set. 

One of the main purpose of defining the new split of PDBBind is to prevent data leakage between training, validation, and test data. To understand whether the new split has solved the data leakage issue, the maximum similarity for proteins and ligands between the training

\noindent
validation, and test data under the EquiBind split and LP-PDBBind, and maximum similarity between the general set and core set under the original PDBBind split are summarized in Figure \ref{fig:pdbbind-data-stats}(d-g). The original PDBBind split has significant protein and ligand overlap between the general and core sets, as can be seen in the sharp peak at similarity of 1.0 in Figure \ref{fig:pdbbind-data-stats}(d,f). The same level of similarity between the refined and core sets in the original PDBBind split are shown in Supplementary Figure S2. Since many ML models use the PDBBind core set as the test dataset without carefully excluding similar data from their respective training dataset, the  performance of these models reported may be overly optimistic and do not reflect their true generalizability and needs to be reevaluated.

The Equibind split is a significant step forward in reorganizing the PDBBind data more reasonably. Its time-based cutoff in defining the test dataset decreased the chance of data leakage, but still did not eliminate the possibility of highly-similar data that occurs in both train and test set. By comparison, LP-PDBBind minimizes overlap between data in the train and test set by design, keeping the maximum sequence similarty between any protein in the test set and any protein in the train set below 0.5, and the maximum ligand similarity below 0.99 (Figure \ref{fig:pdbbind-data-stats}(d,f)). Therefore, results for a model trained with the LP-PDBBind train set and evaluated with LP-PDBBind test set should better reflect the performance of the model when applied to a new protein-ligand complex that may be very different than data used for training the model.

Given that proteins with low sequence similarity might still be homologous and contain binding pockets that share similar interaction characteristics\cite{generalization}, it is essential to double check on the similarities of protein-ligand interaction patterns for complexes in this new split of PDBBind dataset. The proteo-chemometrics based interaction fingerprints \cite{fingerprint} were used to evaluate the interaction pattern similarities. Figure \ref{fig:pdbbind-data-stats}h shows the distribution of the maximum interaction fingerprint similarities for each complex in the test dataset with all complex structures in the training dataset under both the new LP-PDBBind split and the time-based Equibind split, and also the comparison between data in PDBBind core set and general set. LP-PDBBind has also eliminated data with high fingerprint similarity between train and test, which means the data with low protein sequence or ligand similarity but high interaction similarity is not an issue for the LP-PDBBind dataset, and the performance of any scoring function trained with LP-PDBBind train set and evaluated with the test set should be a realistic reflection of the model performance even when the protein-ligand interaction patterns are quite different from the complexes used for training.

To help with training more transferable models, we have also defined the validation set to be equally different than the train set as the test set. Figure \ref{fig:pdbbind-data-stats}(e,g,i) illustrates the maximum protein similarities, maximum ligand similarities and maximum interaction fingerprint similarities between the validation and training sets for the Equibind split and LP-PDBBind, respectively. The similarity between validation and training sets under LP-PDBBind is also well controlled so that the validation set can be used to select the most transferable model or hyperparameters when training the model. By comparison, the validation set in Equibind split is too similar to the train set, hence overfitting will not be effectively captured when monitoring model performance on the validation set. As is provided in Supplementary Figure S3, both the Equibind and LP-PDBBind splits of PDBBind, the validation and test set have a wide range of (dis)/similarities on proteins, ligands and their interactions. Therefore using the validation set for model selection will not lead to overfitting and thus increase transferability to the test set.

\subsection{Evaluating Retrained Models with LP-PDBBind}
After the AutoDock Vina, IGN, RF-Score and DeepDTA models were retrained using LP-PDBBind, the performance were first compared on the LP-PDBBind test dataset filtered to CL2 for better reliability of data. Table \ref{tab:LP-PDBBind-results} provides the mean and standard deviations on the root mean square error (RMSE) of the binding affinity prediction ($\Delta G_{bind}$) on the training, validation, and test data for the models retrained with LP-PDBBind from 3 different model initializations, and comparing it to the original models and their performance on the LP-PDBBind test data. Supplementary Figure S4 shows the scatter plots of the best model and reports the correlation coefficient between predicted and experimental binding affinities. 

\vspace{5mm}

\begin{table}[h]
    \centering
    \caption{\textit{Training and test errors for the original and retrained models using LP-PDBBind.} Results for AutoDock Vina, IGN, RF-Score, and DeepDTA in terms of root mean square error (RMSE) in kcal/mol. Reported standard deviations were calculated with 3 different random initializations of model weights during training.} 
    \begin{tabular}{lccccc}
\hline\hline
\textbf{Model} & \multicolumn{1}{c}{\textbf{RMSE Original}} & & \multicolumn{3}{c}{\textbf{RMSE Retrained}} \\ 
& test & &train & validation &test \\\hline
AutoDock Vina &2.85& &2.42 $\pm$ 0 &2.29 $\pm$ 0 &2.56 $\pm$ 0\\
IGN &1.82 &&1.65 $\pm$ 0.07 & 2.00$\pm$ 0.03 &2.16$\pm$ 0.13 \\
RF-Score &1.89& &0.68 $\pm$ 0.003 &2.14 $\pm$ 0.01 & 2.10 $\pm$ 0.003\\
DeepDTA &1.34 &&1.41$\pm$ 0.11 & 2.07$\pm$ 0.02& 2.29$\pm$ 0.04\\                          
        \hline
    \end{tabular}
\label{tab:LP-PDBBind-results}
\end{table}

Due to the data leakage issue, performance of the original models on the LP-PDBBind test data are over-estimated for the MLSFs. By comparison, AutoDock Vina due to its small number of trainable parameters does not suffer from the data leakage issue, and has achieved lower RMSE after retraining. Among the MLSF models, IGN has the smallest generalization gap, and is also the best performing model when evaluated using the LP-PDBBind test set within uncertainties. Since random forest models can almost perfectly fit training data \cite{PERT,overfitting_or_perfect_fitting}, the RF-Score model has exceptional performance on the LP-PDBBind training dataset, but its validation and test performance is on par with the IGN model. DeepDTA also performs quite well on the training dataset, but exhibits a large generalization gap with respect to the validation dataset, and has slightly higher RMSE on the test dataset than the other two MLSFs. Overall the original MLSF models have seen some of the LP-PDBBind test proteins and ligands in their training, and thus they appear to perform better than they actually do when data leakage is controlled for using the LP-PDBBind data.

Table \ref{tab:bdb-results} summarizes the scoring performance of the original and retrained models on the independent BDB2020+ benchmark dataset; the corresponding scatter plots for the original and retrained models are provided in Supplementary Figure S5. We see that upon retraining, all four models have achieved meaningful improvements on the BDB2020+ benchmark set. In terms of RMSE, AutoDock Vina decreased by 1.2 kcal/mol while the other three MLSFs decreased by 0.2-0.3 kcal/mol. The changes in the correlation coefficients are more profound: both IGN and RF-Score have achieved absolute correlation coefficients better than 0.5, and improvements for AutoDock Vina and DeepDTA are also clear. Overall, IGN and RFScore have improved more significantly than the other two, especially IGN that has achieved 1.38 $\pm$ 0.09 kcal/mol in RMSE and 0.54$\pm$ 0.04 in correlation coefficient when evaluated on the BDB2020+ dataset. We attribute the improvements on the model performances to our resplit of the PDBBind dataset because the training dataset becomes more diverse and representative, and that helps the SFs to find more transferable features and make more accurate predictions.

\vspace{5mm}

\begin{table}[h]
    \centering
    \caption{\textit{Performance comparisons on the BDB2020+ benchmark set.} Root mean square error (RMSE) and Pearson correlation coefficient ($R$) for different models before and after retraining using LP-PDBBind. Reported standard deviations were calculated with 3 different random initializations of model weights during training.}

    \begin{tabular}{lcccccc}
        \hline\hline

\textbf {Model } & \multicolumn{3}{c}{\textbf{RMSE (kcal/mol)}} & \multicolumn{2}{c}{\textbf{\textit{R}}} \\ 
& original & retrained & difference & original & retrained \\\hline
AutoDock Vina&3.31&2.10 $\pm$ 0 &-37\%&0.23&0.29$\pm$0\\
IGN&1.62&1.38$\pm$ 0.09&-9$\sim$20\%&0.41&0.54$\pm$0.04\\
RF-Score&1.80&1.61 $\pm$ 0.008&-10$\sim$11\%&0.36&0.51$\pm$ 0.008\\
DeepDTA&1.98&1.72$\pm$0.10 &-9$\sim$19\%&0.18&0.26$\pm$0.07\\
                           
        \hline
    \end{tabular}

\label{tab:bdb-results}
\end{table}

\begin{figure}[H]
\begin{center}
\includegraphics[width=0.9\textwidth]{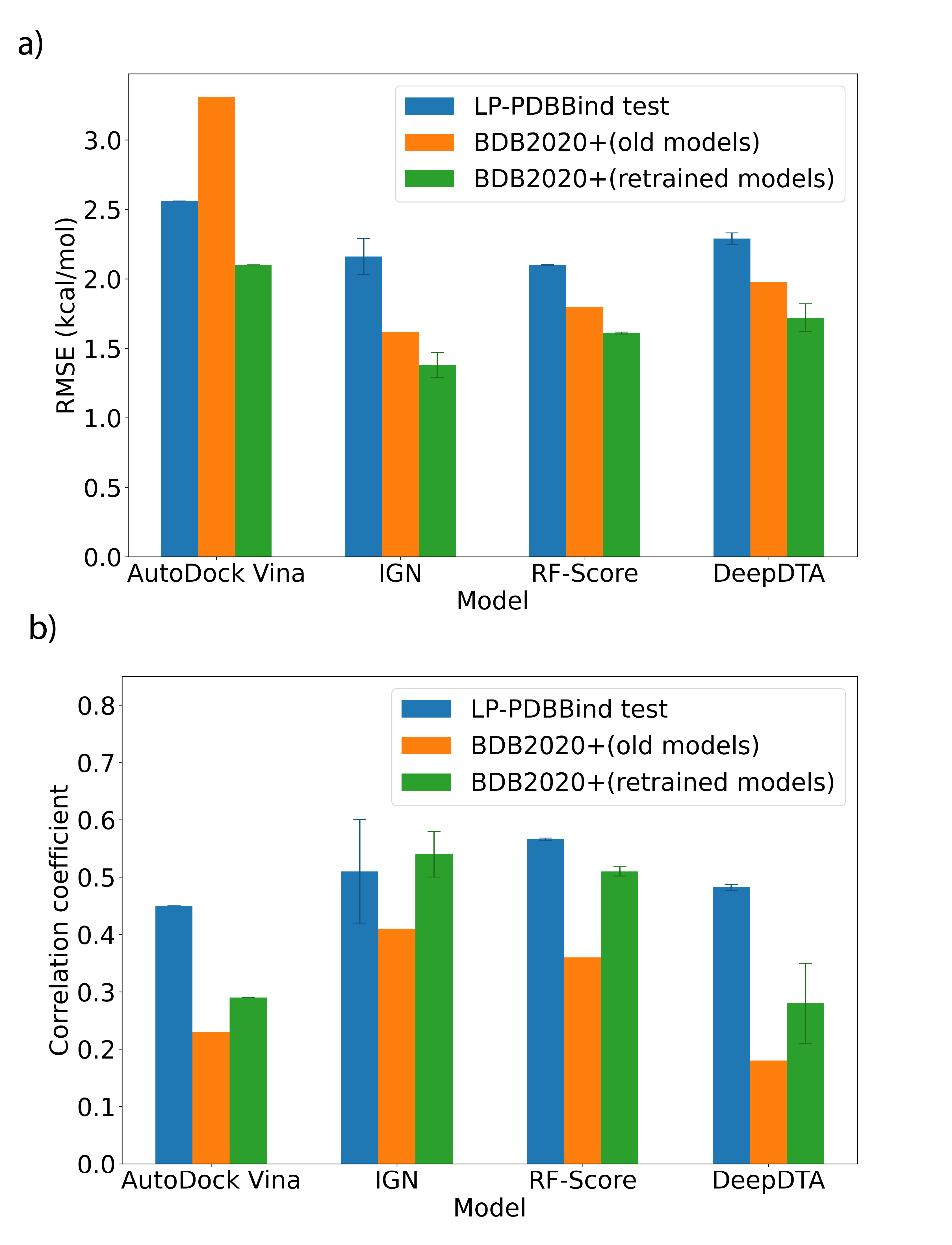}
\end{center}
\caption{\textit{Performance comparisons using different models and different benchmark datasets.} (a) Comparison on the root mean square error (RMSE) for different models. Lower is better. Blue bars indicate RMSEs on the LP-PDBBind test dataset using retrained models, orange bars indicate RMSEs for the models without retraining using LP-PDBBind, and green bars indicate RMSEs for the models retrained using LP-PDBBind. (b) Comparison on the Pearson correlation coefficient ($R$) for different models.}
\label{fig:bdb-results}
\end{figure}

Figure \ref{fig:bdb-results} also compares the performance of the four models on BDB2020+ with the results evaluated on the LP-PDBBind test dataset. Interestingly, we find that all models have achieved lower RMSE on the BDB2020+ dataset but also lower Pearson correlation coefficients, $R$, other than IGN that has similar correlation coefficients between LP-PDBBind test dataset and BDB2020+ benchmark set. This seemingly contradictory result is actually reasonable due to the dataset distribution differences of the two evaluation benchmarks. The LP-PDBBind test dataset contains much more data than BDB2020+, and also spans a wider range of binding affinity values. The measured $-\log(K_d)$ values in the LP-PDBBind test dataset ranges from 0 to 12 (i.e. 12 orders of magnitude), but the BDB2020+ dataset only 

\noindent
ranges from 4 to 10. Given that extreme predictions from a robust ML model is unlikely, a narrower range of binding affinities means the overall error will be smaller. However, it also poses challenge for successfully differentiating the nuances between more clustered data points, and therefore it is also more difficult to achieve higher correlation coefficients.

Nevertheless, we see that the relative rankings of the four methods are consistent between different evaluation benchmarks and different metrics. We find that IGN and RF-Score perform consistently better than AutoDock Vina and DeepDTA when trained on the new split. In terms of a generalization gap to new data, we also see the correlation differences are the smallest for IGN and RF-Score. These results are consistent with the overall performances of the models, and provide evidence that modern MLSFs can indeed surpass CSFs such as AutoDock Vina, even when protein and ligand similarities are low. While DeepDTA as a MLSF is an exception to this conclusion, we suggest that it is because it does not explicitly take molecular geometries as input, and it would perhaps explain why it does not achieve the same level of accuracy as MLSF or CSF models relying on 3-dimensional information of protein-ligand complexes.

\subsection{Evaluating the Ranking Capabilities of the Retrained Models}
The improvement in scoring power, despite being significant, still does not fully reflect the ranking capabilities of the SFs which are more important in deciding which compound to prioritize in a real-world drug discovery campaign. Therefore, we have prepared an Mpro dataset and an EGFR dataset to evaluate the ranking accuracy of the SFs before and after retraining using LP-PDBBind.

\begin{figure}[H]
\begin{center}
\includegraphics[width=0.85\textwidth]{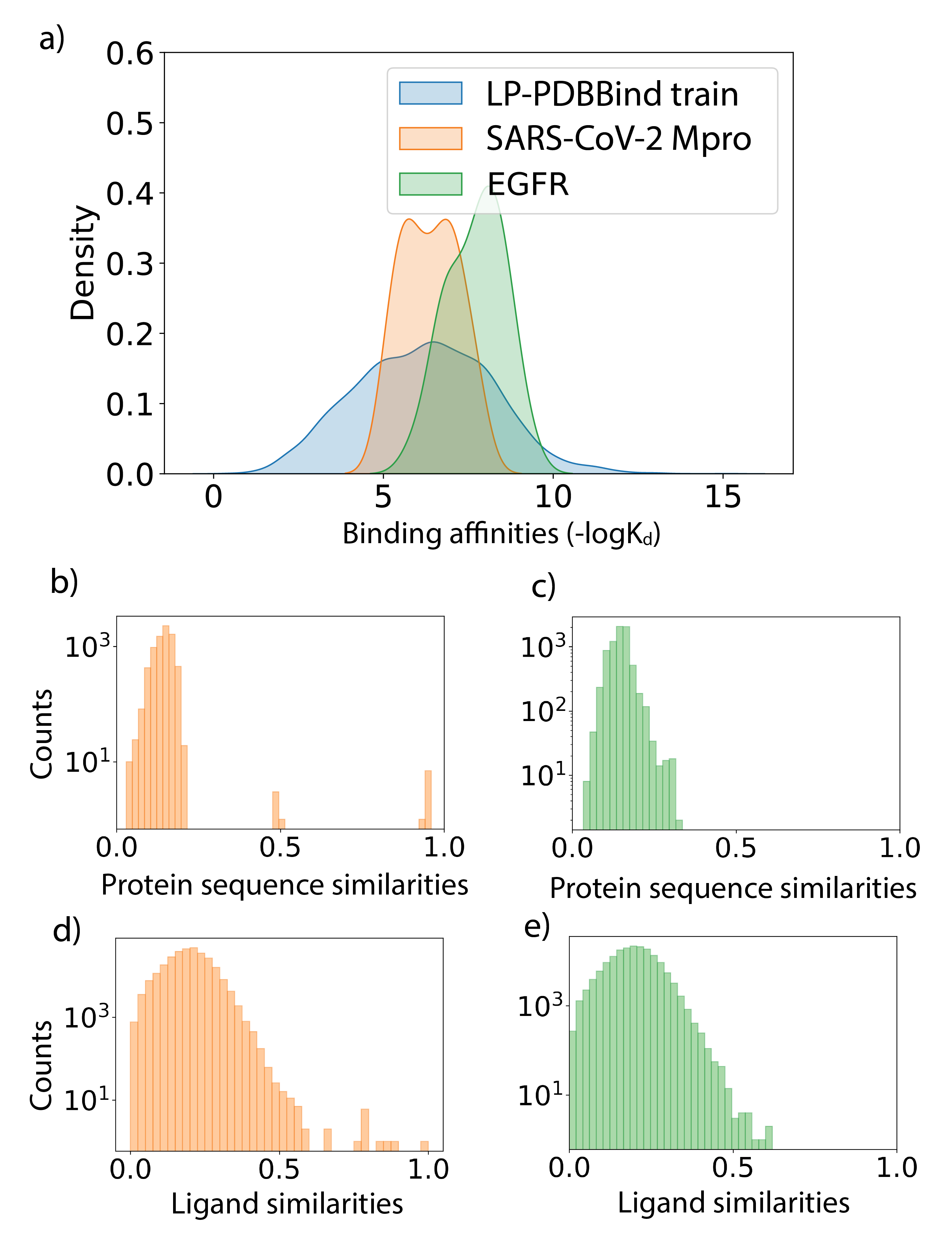}
\end{center}
\caption{\textit{Data statistics for the SARS-CoV-2 main protease (Mpro) benchmark set and epidermal growth factor receptor (EGFR) benchmark set.} (a) Distributions of the binding affinity data ($-\log K_d$) in the LP-PDBBind train dataset in blue, Mpro benchmark set in orange and EGFR set in green. (b-c) Distributions of protein sequence similarities between the Mpro protein (b) and EGFR protein (c) with proteins in the LP-PDBBind train dataset. (d-e) Distributions of ligand fingerprint similarities between molecules in the Mpro benchmark set (d) and EGFR benchmark set (e) with ligands in the LP-PDBBind train dataset.}
\label{fig:ranking-properties}
\end{figure}

Figure \ref{fig:ranking-properties} shows the distributions of experimental binding affinities and protein and ligand similarities when comparing the LP-PDBBind training data with the Mpro and EGFR datasets. The binding affinities for these two systems are found to be in a narrower range than the LP-PDBBind training data. While the average binding affinity of the Mpro dataset is roughly in line with the LP-PDBBind training dataset, the ligands in the EGFR dataset 

\noindent
have an overall tendency to bind stronger to their targets. This shift in the average indicates that the EGFR is more "out-of-distribution" than the Mpro dataset. By calculating the Kullback-Leibler divergence (KL-divergence)\cite{kldiv} with the kernel density estimations between binding affinities in the LP-PDBBind train dataset and the two target-specific datasets, we found that the KL-divergence between LP-PDBBind train dataset and SARS-COV-2 Mpro dataset is 3.45, and the KL-divergence between LP-PDBBind train dataset and EGFR dataset is 4.46, also validating the EFGR dataset has more dissimilar binding affinities to the training dataset.  Figure \ref{fig:ranking-properties}b-e further show that the protein and sequence similarities of the two evaluation datasets are overall very dissimilar to the LP-PDBBind training dataset, with the exception of a small fraction of protein and ligand sequence similarity attributable to the SARS-CoV-1 Mpro training entry PDB ID: 3V3M\cite{pdb_3v3m} for the Mpro dataset. Hence the Mpro and EFGR evaluations reflect two representative scenarios of using the SFs on a similar or different protein-ligand system than what wasincluded in the PDBBind dataset.

Table \ref{tab:mproefgr-results} summarizes the RMSE, Pearson correlation coefficients ($R$) and Spearmann correlation coefficients ($R_S$ or ranking power) for the data in the Mpro and EGFR evaluation sets using both the original models and the models retrained using LP-PDBBind. The prediction scatter plots for the Mpro benchmark dataset and EGFR benchmark dataset are provided in Supplementary Figures S6 and S7. For the Mpro dataset, all four models exhibit some modest improvement in terms of RMSE. AutoDock Vina, RF-Score and DeepDTA have all achieved higher correlations in terms of $R$ and $R_S$. The correlation on the retrained IGN model has slightly decreased but is still similar to the original values. Taken all metrics into account, AutoDock Vina is the best performing SF for the Mpro dataset, although the differentiation among all models is not large.

\begin{table}[h]
    \centering
    \caption{Performance comparisons on the SARS-CoV-2 main protease (Mpro) and epidermal growth factor receptor (EGFR) benchmark set in terms of root mean square error (RMSE) in kcal/mol, Pearson correlation coefficient ($R$) and Spearmann correlation coefficient ($R_S$) for different models before and after retraining using LP-PDBBind. The retrained model with the best performance on the BDB2020+ benchmark dataset was used to predict binding affinities in the Mpro and EGFR dataset.}
    \begin{tabular}{lcccccc}
        \hline\hline
\textbf {Mpro Model } & \multicolumn{2}{c}{\textbf{RMSE (kcal/mol)}} & \multicolumn{2}{c}{\textbf{\textit{R}}} &\multicolumn{2}{c}{$\boldsymbol{R_S}$} \\ 
& original & retrained  & original & retrained & original & retrained  \\\hline
AutoDock Vina&1.20&1.17&0.55&0.66&0.51&0.68\\
IGN&1.86&1.44&0.64&0.61&0.69&0.65\\
RF-Score&2.06&1.64&0.43&0.52&0.47&0.58\\
DeepDTA&1.18&0.88&0.59&0.64&0.60&0.65\\ 
        \hline\hline
\textbf {EFGR Model } & \multicolumn{2}{c}{\textbf{RMSE (kcal/mol)}} & \multicolumn{2}{c}{\textbf{\textit{R}}} &\multicolumn{2}{c}{$\boldsymbol{R_S}$} \\ 
& original & retrained  & original & retrained & original & retrained  \\\hline
AutoDock Vina&3.11&1.59&0.25&0.38&0.21&0.36\\
IGN&1.06&0.96&0.36&0.65&0.17&0.62\\
RF-Score&1.57&0.97&-0.15&0.52&-0.18&0.45\\
DeepDTA&1.22&1.05&0.23&0.44&0.20&0.43\\         
        \hline
    \end{tabular}

\label{tab:mproefgr-results}
\end{table}

However, the results on the EGFR dataset exhibit much larger variations for the different models, and the changes are significant for all the SFs. There is a dramatic decrease by \char`\~ 50\% in RMSE for AutoDock Vina, and both $R$ and $R_S$ have increased significantly. IGN has a slight decrease in RMSE but the correlation coefficients have improved remarkably from 0.36 to 0.65 in Pearson correlation coefficient and from 0.17 to 0.62 in in Spearmann correlation coefficient. These values are the highest among all retrained models, and do not differ much with the Mpro results and thus showing good generalizability. RF-Score has also benefitted from retraining quite significantly, with a 0.6 kcal/mol decrease in RMSE, and the correlation coefficients have improved from -0.15 to 0.52 in terms of $R$ and from -0.18 to 0.45 in terms of $R_S$. Finally, DeepDTA has also seen improvements upon retraining, but the ranking capability of the retrained DeepDTA model is still inferior to the other two MLSFs. Overall, the performance rankings of the four SFs are still consistent with that obtained from the LP-PDBBind test set benchmark and the BDB 2020+ benchmark. But the EFGR benchmark emphasizes that new applications will benefit from better generalizability of the newly retrained models, especially for IGN.

\subsection{Rescoring Docked Poses and Pose Selection using the Retrained Scoring Functions}
Cocrystal structures are usually unavailable during the early stages of the drug discovery process. This it is important to quantify the binding affinity estimations from the retrained SFs using predicted structures, such as the complex structures obtained from molecular docking. Using the BDB2020+ structures redocked with AutoDock Vina and the original Vina scoring function, we have evaluated the accuracies of the retrained AutoDock Vina, IGN and RF-Score as a rescoring method, and the results are summarized in Table \ref{tab:redock-results}; DeepDTA was not compared because its predictions do not rely on the exact 3-dimensional structures of the complex. 

Following a recent study that suggests a modified set of empirical parameters for AutoDock Vina that improves its ranking capabilities, we have also evaluated its performance on the redocked BDB2020+ benchmark set.\cite{vina_new_weights} To better separate pose selection capabilities from scoring capabilities, we report metrics on both the redocked structures with the best docking scores, and the redocked poses that have the lowest RMSDs to the cocrystal structures.

\begin{table}[h]
    \centering
    \caption{\textit{Performance comparisons on scoring with redocked structures of the BDB2020+ benchmark.} Root mean square error (RMSE) and Pearson correlation coefficient (R) for different models before and after retraining using LP-PDBBind. The retrained model with the best performance on the BDB2020+ benchmark dataset was used in this benchmark. AutoDock Vina with the newly suggested weights in Ref. \citenum{vina_new_weights} was also added for comparison. }
    \begin{tabular}{lccccc}
        \hline\hline
\multirow{2}{8cm}{\textbf {\shortstack[l]{Redocked Structures\\ (Best Score) }}} & \multicolumn{2}{c}{\textbf{RMSE (kcal/mol)}} & \multicolumn{2}{c}{\textbf{\textit{R}}}  \\ 
& original & retrained  & original & retrained   \\\hline
AutoDock Vina&2.12&1.99&0.22&0.26\\
AutoDock Vina with Modified Weights (Ref. \citenum{vina_new_weights})&3.35&/&0.25&/&\\
RF-Score&1.78&1.56&0.34&0.49\\
IGN&1.66&1.42&0.33&0.46\\ 
        \hline\hline
\multirow{2}{8cm}{\textbf {\shortstack[l]{Redocked Structures \\(Closest to Cocrystal)} }} & \multicolumn{2}{c}{\textbf{RMSE (kcal/mol)}} & \multicolumn{2}{c}{\textbf{\textit{R}}}  \\ 
& original & retrained  & original & retrained   \\\hline
AutoDock Vina&2.18&2.02&0.13&0.22\\
AutoDock Vina with Modified Weights (Ref. \citenum{vina_new_weights})&3.40&/&0.17&/&\\
RF-Score&1.83&1.61&0.27&0.46\\
IGN&1.65&1.43&0.34&0.43\\         
        \hline
    \end{tabular}

\label{tab:redock-results}
\end{table}

Again, we observed a consistent improvement over the original models after retraining the SFs using LP-PDBBind, which means even though the models were trained using cocrystal structures, they could be used to rescore docked poses and achieve higher accuracies when compared with experimental binding affinities. In the top-scored redocked structure in BDB2020+ benchmark set, AutoDock Vina has improved from 2.12 to 1.99 kcal/mol in RMSE and the correlation coefficients improved from 0.22 to 0.26. Compared to the suggested new weights in Ref.\citenum{vina_new_weights}, the AutoDock Vina 
scoring function retrained with LP-PDBBind still has better scoring and ranking capabilities. The improvements on the retrained RF-Score and IGN models are also distinct, especially in the correlation coefficients that have improved from 0.34 to 0.49 for RF-Score and from 0.33 to 0.46 for IGN. The performance on the redocked structures that are closest to the known cocrystal structure eliminates the influence of bad docked poses and purely reflects the scoring and ranking capabilities of the retrained models on correctly docked structures. The performance was similar to that evaluated on the best-scored poses, further showing that the improvements obtained by retraining the models using cocrystal structures are transferable to docked structures.

However, we would like to emphasize that the newly trained Vina scoring function is not intended for docking, because a scoring function designed for docking should be trained with high energy decoy poses which are missing in the PDBBind dataset. Consequently, the repulsion parameter in our retrained Vina scoring function (0.014212) is much smaller than the original value (0.840) and the scoring function cannot differentiate bad poses that have atoms too close to each other. Nevertheless, the improvements shown for scoring power and ranking power are meaningful, because the virtual
screening process can be broken into multiple steps, and we could use the original Vina SF for docking and the retrained one to rescore top poses to achieve better accuracy. This is similar to how IGN and RF-Score are used for example, and thus the retrained AutoDock-vina can also be used the same way.

\section{Discussion and Conclusion}
The area of computational drug discovery relies on generalizable scoring functions that have robust scoring and ranking power of binding affinities of ligand-protein complexes. However due to the lack of independently built datasets that test the true generalizability of the SFs, it is hard to differentiate among the plethora of many models which have been trained and tested on the original split of the PDBBind dataset. As we have shown, there is too much overlap between the PDBBind general and refined data used for training with the core subset, leading to the possibility of inflated performance metrics that in turn lead to false confidence in how such models will perform on new protein-ligand complexes.\cite{buttenschoen2023posebusters} 

In order to reduce data similarity between training, validation, and test data of the PDBBind dataset, we have developed LP-PDBBind using an iterative process to select most similar data first into the test set, and then validation set, so that the final training dataset has low similarity with validation or test dataset. Furthermore, to provide a benchmark dataset truly independent of PDBBind, we have compiled the BDB2020+ dataset derived from the BindingDB database deposited after 2020 and ending in 2023 to minimize time-cutoff based data leakage, and further ensuring that there is no overlap with PDBBind. Additionally, we also tested ranking power through construction of the Mpro and EFGR ligand-protein complex series. The SARS-CoV-2 Mpro protein has high similarity counterparts in the PDBBind dataset, and the other series involving EGFR does not have anything similar in the training dataset of PDBBind. These new data should also be useful in future evaluation and/or finetuning of any scoring function.

In this work we utilized the new split of PDBBind to retrain AutoDock Vina, IGN, RF-Score and DeepDTA and compared the old models with the retrained models on the LP-PDBBind test set, as well as the fully independent BDB2020+ data, Mpro series, and EGFR series. We have demonstrated that using a different splitting of the same dataset leads to significant performance improvements, and the improvements are transferable from using co-crystal structures as input to using docked structures as input. The comparison between the different benchmark results for the SFs can be well explained by the "difficulty level" of these datasets, and provide insights about the generalizability of various models. We found that well performing models also have more stable ranking results for different ligands towards a protein target. When the target system is similar to data included in the training dataset, the differentiation between models are not obvious. However, when the protein target is not similar to anything in the training dataset, we found that different SFs demonstrate quite different generalization capabilities; the IGN model retrained with LP-PDBBind is recommended due to its reliable good scoring and ranking power.

However, since the PDBBind dataset does not contain decoy structures that are important to allow the models to learn characteristics of true complex structures, the retrained SFs are not expected to have better docking power for recognizing native ligand bound poses. The usefulness of the retrained SF relies on their capability to rescore docked poses and achieve better accuracy when compared to experimental binding affinity measurements. Meanwhile, it would be worthwhile to generate decoy structures for the BDB2020+ dataset to better benchmark the docking and screening powers of the SFs independent of the PDBBind dataset. In summary, we have created more transferable SFs, and we hope reporting evaluation metrics on the BDB2020+ dataset can also become a common practice for future SFs.

\begin{acknowledgement}
We sincerely thank Prof. Michael Gilson for assistance with the BDBind data set. This work was supported by National Institute of Allergy and Infectious Disease grant U19-AI171954. This research used computational resources of the National Energy Research Scientific Computing, a DOE Office of Science User Facility supported by the Office of Science of the U.S. Department of Energy under Contract No. DE-AC02-05CH11231. 
\end{acknowledgement}

\section{Data Availability and Implementation}
All the code for building the BDB2020+ dataset, the prepared dataset files in the .csv format, and python scripts for retraining AutoDock Vina, IGN, RFScore and DeepDTA models are provided in a public accessible GitHub repository: https://github.com/THGLab/LP-PDBBind
\begin{suppinfo}
Computational procedures and characterization of the PDBBind, EquiBind, and LP-PDBBind data sets, and results for original and retrained scoring functions. 
\end{suppinfo}

\begin{figure}[H]
\begin{center}
\includegraphics[width=0.99\textwidth]{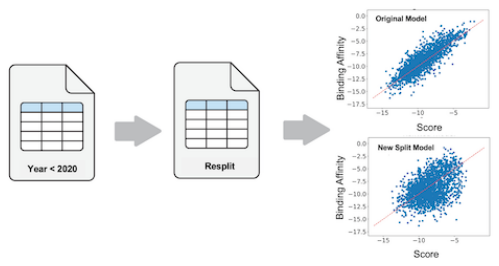}
\vspace{-7mm}
\end{center}
\caption{TOC graphic }

\label{fig:toc}
\end{figure}

\bibliography{references}

@article{autodock_vina,
  title={AutoDock Vina: improving the speed and accuracy of docking with a new scoring function, efficient optimization, and multithreading},
  author={Trott, Oleg and Olson, Arthur J},
  journal={J. Comp. Chem. },
  volume={31},
  number={2},
  pages={455--461},
  year={2010},
  publisher={Wiley Online Library}
}

@misc{li2023LP,
      title={Leak Proof PDBBind: A Reorganized Dataset of Protein-Ligand Complexes for More Generalizable Binding Affinity Prediction}, 
      author={Jie Li and Xingyi Guan and Oufan Zhang and Kunyang Sun and Yingze Wang and Dorian Bagni and Teresa Head-Gordon},
      year={2023},
      eprint={2308.09639},
      archivePrefix={arXiv},
      primaryClass={physics.bio-ph},
      url={https://arxiv.org/abs/2308.09639}, 
}

@article{Wang2025,
   author = {Wang, Yingze and Sun, Kunyang and Li, Jie and Guan, Xingyi and Zhang, Oufan and Bagni, Dorian and Zhang, Yang and Carlson, Heather A. and Head-Gordon, Teresa},
   title = {A workflow to create a high-quality protein–ligand binding dataset for training, validation, and prediction tasks},
   journal = {Digital Discovery},
   volume = {4},
   number = {5},
   pages = {1209-1220},
   DOI = {10.1039/D4DD00357H},
   url = {http://dx.doi.org/10.1039/D4DD00357H},
   year = {2025},
   type = {Journal Article}
}

@article{Datasail2025,
   author = {Joeres, Roman and Blumenthal, David B. and Kalinina, Olga V.},
   title = {Data splitting to avoid information leakage with DataSAIL},
   journal = {Nature Communications},
   volume = {16},
   number = {1},
   pages = {3337},
   ISSN = {2041-1723},
   DOI = {10.1038/s41467-025-58606-8},
   url = {https://doi.org/10.1038/s41467-025-58606-8},
   year = {2025},
   type = {Journal Article}
}

@article{Graber2025,
   author = {Graber, David and Stockinger, Peter and Meyer, Fabian and Mishra, Siddhartha and Horn, Claus and Buller, Rebecca},
   title = {Resolving data bias improves generalization in binding affinity prediction},
   journal = {Nature Machine Intelligence},
   volume = {7},
   number = {10},
   pages = {1713-1725},
   ISSN = {2522-5839},
   DOI = {10.1038/s42256-025-01124-5},
   url = {https://doi.org/10.1038/s42256-025-01124-5},
   year = {2025},
   type = {Journal Article}
}

@article{autodock4,
  title={AutoDock4 and AutoDockTools4: Automated docking with selective receptor flexibility},
  author={Morris, Garrett M and Huey, Ruth and Lindstrom, William and Sanner, Michel F and Belew, Richard K and Goodsell, David S and Olson, Arthur J},
  journal={J. Comp. Chem. },
  volume={30},
  number={16},
  pages={2785--2791},
  year={2009},
  publisher={Wiley Online Library}
}

@article{pdbbind,
  title={PDB-wide collection of binding data: current status of the PDBbind database},
  author={Liu, Zhihai and Li, Yan and Han, Li and Li, Jie and Liu, Jie and Zhao, Zhixiong and Nie, Wei and Liu, Yuchen and Wang, Renxiao},
  journal={Bioinformatics},
  volume={31},
  number={3},
  pages={405--412},
  year={2015},
  publisher={Oxford University Press}
}

@article{tankbind,
  title={Tankbind: Trigonometry-aware neural networks for drug-protein binding structure prediction},
  author={Lu, Wei and Wu, Qifeng and Zhang, Jixian and Rao, Jiahua and Li, Chengtao and Zheng, Shuangjia},
  journal={bioRxiv},
  pages={2022--06},
  year={2022},
  publisher={Cold Spring Harbor Laboratory}
}

@article{rfscore,
  title={A machine learning approach to predicting protein--ligand binding affinity with applications to molecular docking},
  author={Ballester, Pedro J and Mitchell, John BO},
  journal={Bioinformatics},
  volume={26},
  number={9},
  pages={1169--1175},
  year={2010},
  publisher={Oxford University Press}
}

@article{IGN,
  title={Interactiongraphnet: A novel and efficient deep graph representation learning framework for accurate protein--ligand interaction predictions},
  author={JJiang, Dejun and Hsieh, Chang-Yu and Wu, Zhenxing and Kang, Yu and Wang, Jike and Wang, Ercheng and Liao, Ben and Shen, Chao and Xu, Lei and Wu, Jian and Cao, Dongsheng and Hou, Tingjun},
  journal={J. Med. Chem. },
  volume={64},
  number={24},
  pages={18209--18232},
  year={2021},
  publisher={ACS Publications}
}

@article{pignet,
  title={PIGNet: a physics-informed deep learning model toward generalized drug--target interaction predictions},
  author={Moon, Seokhyun and Zhung, Wonho and Yang, Soojung and Lim, Jaechang and Kim, Woo Youn},
  journal={Chemical Science},
  volume={13},
  number={13},
  pages={3661--3673},
  year={2022},
  publisher={Royal Society of Chemistry}
}

@misc{buttenschoen2023posebusters,
      title={PoseBusters: AI-based docking methods fail to generate physically valid poses or generalise to novel sequences}, 
      author={Martin Buttenschoen and Garrett M. Morris and Charlotte M. Deane},
      year={2023},
      eprint={2308.05777},
      archivePrefix={arXiv},
      primaryClass={q-bio.QM}
}

@article{LeePing,
   author = {Wang, Lee-Ping and Martinez, Todd J. and Pande, Vijay S.},
   title = {Building Force Fields: An Automatic, Systematic, and Reproducible Approach},
   journal = {J. Phys. Chem. Lett.},
   volume = {5},
   number = {11},
   pages = {1885-1891},
   DOI = {10.1021/jz500737m},
   url = {https://doi.org/10.1021/jz500737m},
   year = {2014},
   type = {Journal Article}
}

@article{Haghighatlari2020,
   author = {Haghighatlari, Mojtaba and Li, Jie and Heidar-Zadeh, Farnaz and Liu, Yuchen and Guan, Xingyi and Head-Gordon, Teresa},
   title = {Learning to Make Chemical Predictions: The Interplay of Feature Representation, Data, and Machine Learning Methods},
   journal = {Chem},
   volume = {6},
   number = {7},
   pages = {1527-1542},
   DOI = {https://doi.org/10.1016/j.chempr.2020.05.014},
   url = {https://doi.org/10.1016/j.chempr.2020.05.014},
   year = {2020},
   type = {Journal Article}
}

@article{deepDTA,
   author = {Ozturk, Hakime and Ozgur, Arzucan and Ozkirimli, Elif},
   title = {DeepDTA: deep drug–target binding affinity prediction},
   journal = {Bioinformatics},
   volume = {34},
   number = {17},
   pages = {i821-i829},
   ISSN = {1367-4803},
   DOI = {10.1093/bioinformatics/bty593},
   url = {https://doi.org/10.1093/bioinformatics/bty593},
   year = {2018},
   type = {Journal Article}
}

@article{deeplearning_vs_traditional_docking,
  title={Do deep learning models really outperform traditional approaches in molecular docking?},
  author={Yu, Yuejiang and Lu, Shuqi and Gao, Zhifeng and Zheng, Hang and Ke, Guolin},
  journal={arXiv preprint arXiv:2302.07134},
  year={2023}
}

@article{rdkit,
  title={RDKit: A software suite for cheminformatics, computational chemistry, and predictive modeling},
  author={Landrum, Greg},
  journal={Greg Landrum},
  volume={8},
  year={2013}
}

@article{needleman_wunsch,
  title={A general method applicable to the search for similarities in the amino acid sequence of two proteins},
  author={Needleman, Saul B and Wunsch, Christian D},
  journal={J. Mol. Bio.},
  volume={48},
  number={3},
  pages={443--453},
  year={1970},
  publisher={Elsevier}
}

@article{guo2022covbinderinpdb,
  title={CovBinderInPDB: A Structure-Based Covalent Binder Database},
  author={Guo, Xiao-Kang and Zhang, Yingkai},
  journal={J. Chem. Info. Model.},
  volume={62},
  number={23},
  pages={6057--6068},
  year={2022},
  publisher={ACS Publications}
}

@article{glide1,
  title={Glide: a new approach for rapid, accurate docking and scoring. 1. Method and assessment of docking accuracy},
  author={Friesner, Richard A. and Banks, Jay L. and Murphy, Robert B. and Halgren, Thomas A. and Klicic, Jasna J. and Mainz, Daniel T. and Repasky, Matthew P. and Knoll, Eric H. and Shelley, Mee and Perry, Jason K. and Shaw, David E. and Francis, Perry and Shenkin, Peter S.},
  journal={J. Med. Chem. },
  volume={47},
  number={7},
  pages={1739--1749},
  year={2004},
  publisher={ACS Publications}
}

@article{glide2,
  title={Extra precision glide: Docking and scoring incorporating a model of hydrophobic enclosure for protein- ligand complexes},
  author={Friesner, Richard A and Murphy, Robert B and Repasky, Matthew P and Frye, Leah L and Greenwood, Jeremy R and Halgren, Thomas A and Sanschagrin, Paul C and Mainz, Daniel T},
  journal={J. Med. Chem. },
  volume={49},
  number={21},
  pages={6177--6196},
  year={2006},
  publisher={ACS Publications}
}

@article{xscore,
  title={Further development and validation of empirical scoring functions for structure-based binding affinity prediction},
  author={Wang, Renxiao and Lai, Luhua and Wang, Shaomeng},
  journal={Journal of computer-aided molecular design},
  volume={16},
  pages={11--26},
  year={2002},
  publisher={Springer}
}

@article{OpenFF1,
   author = {Qiu, Y. and Smith, D. G. A. and Boothroyd, S. and Jang, H. and Hahn, D. F. and Wagner, J. and Bannan, C. C. and Gokey, T. and Lim, V. T. and Stern, C. D. and Rizzi, A. and Tjanaka, B. and Tresadern, G. and Lucas, X. and Shirts, M. R. and Gilson, M. K. and Chodera, J. D. and Bayly, C. I. and Mobley, D. L. and Wang, L. P.},
   title = {Development and Benchmarking of Open Force Field v1.0.0-the Parsley Small-Molecule Force Field},
   journal = {J Chem Theory Comput},
   volume = {17},
   number = {10},
   pages = {6262-6280},
   ISSN = {1549-9618 (Print)
1549-9618},
   DOI = {10.1021/acs.jctc.1c00571},
   year = {2021},
   type = {Journal Article}
}

@article{OpenFF2,
   author = {Boothroyd, Simon and Behara, Pavan Kumar and Madin, Owen C. and Hahn, David F. and Jang, Hyesu and Gapsys, Vytautas and Wagner, Jeffrey R. and Horton, Joshua T. and Dotson, David L. and Thompson, Matthew W. and Maat, Jessica and Gokey, Trevor and Wang, Lee-Ping and Cole, Daniel J. and Gilson, Michael K. and Chodera, John D. and Bayly, Christopher I. and Shirts, Michael R. and Mobley, David L.},
   title = {Development and Benchmarking of Open Force Field 2.0.0: The Sage Small Molecule Force Field},
   journal = {J. Chem. Theo. Comp.},
   volume = {19},
   number = {11},
   pages = {3251-3275},
   ISSN = {1549-9618},
   DOI = {10.1021/acs.jctc.3c00039},
   url = {https://doi.org/10.1021/acs.jctc.3c00039},
   year = {2023},
   type = {Journal Article}
}

@article{AD-vina2021,
   author = {Eberhardt, Jerome and Santos-Martins, Diogo and Tillack, Andreas F. and Forli, Stefano},
   title = {AutoDock Vina 1.2.0: New Docking Methods, Expanded Force Field, and Python Bindings},
   journal = {J. Chem. Info. Model.},
   volume = {61},
   number = {8},
   pages = {3891-3898},
   ISSN = {1549-9596},
   DOI = {10.1021/acs.jcim.1c00203},
   url = {https://doi.org/10.1021/acs.jcim.1c00203},
   year = {2021},
   type = {Journal Article}
}

@article{PMF,
  title={PMF scoring revisited},
  author={Muegge, Ingo},
  journal={J. Med. Chem. },
  volume={49},
  number={20},
  pages={5895--5902},
  year={2006},
  publisher={ACS Publications}
}

@article{knowledge_SF,
  title={Knowledge-based scoring function to predict protein-ligand interactions},
  author={Gohlke, Holger and Hendlich, Manfred and Klebe, Gerhard},
  journal={J. Mol. Bio.},
  volume={295},
  number={2},
  pages={337--356},
  year={2000},
  publisher={Elsevier}
}

@article{statistical_potentials,
  title={General and targeted statistical potentials for protein--ligand interactions},
  author={Mooij, Wijnand TM and Verdonk, Marcel L},
  journal={Proteins: Struct., Funct., and Bioinform.},
  volume={61},
  number={2},
  pages={272--287},
  year={2005},
  publisher={Wiley Online Library}
}

@article{rtmscore,
  title={Boosting protein--ligand binding pose prediction and virtual screening based on residue--atom distance likelihood potential and graph transformer},
  author={Shen, Chao and Zhang, Xujun and Deng, Yafeng and Gao, Junbo and Wang, Dong and Xu, Lei and Pan, Peichen and Hou, Tingjun and Kang, Yu},
  journal={J. Med. Chem. },
  volume={65},
  number={15},
  pages={10691--10706},
  year={2022},
  publisher={ACS Publications}
}

@article{holoprot,
  title={Multi-scale representation learning on proteins},
  author={Somnath, Vignesh Ram and Bunne, Charlotte and Krause, Andreas},
  journal={Advances in Neural Information Processing Systems},
  volume={34},
  pages={25244--25255},
  year={2021}
}

@article{accuracy_or_novelty,
  title={Accuracy or novelty: what can we gain from target-specific machine-learning-based scoring functions in virtual screening?},
  author={Shen, Chao and Weng, Gaoqi and Zhang, Xujun and Leung, Elaine Lai-Han and Yao, Xiaojun and Pang, Jinping and Chai, Xin and Li, Dan and Wang, Ercheng and Cao, Dongsheng and Hou, Tingjun},
  journal={Briefings in Bioinformatics},
  volume={22},
  number={5},
  pages={bbaa410},
  year={2021},
  publisher={Oxford University Press}
}

@article{similarity1,
  title={Structural and sequence similarity makes a significant impact on machine-learning-based scoring functions for protein--ligand interactions},
  author={Li, Yang and Yang, Jianyi},
  journal={J. Chem. Info. Model.},
  volume={57},
  number={4},
  pages={1007--1012},
  year={2017},
  publisher={ACS Publications}
}

@article{similarity2,
  title={The impact of protein structure and sequence similarity on the accuracy of machine-learning scoring functions for binding affinity prediction},
  author={Li, Hongjian and Peng, Jiangjun and Leung, Yee and Leung, Kwong-Sak and Wong, Man-Hon and Lu, Gang and Ballester, Pedro J},
  journal={Biomolecules},
  volume={8},
  number={1},
  pages={12},
  year={2018},
  publisher={MDPI}
}

@inproceedings{equibind,
  title={Equibind: Geometric deep learning for drug binding structure prediction},
  author={St{\"a}rk, Hannes and Ganea, Octavian and Pattanaik, Lagnajit and Barzilay, Regina and Jaakkola, Tommi},
  booktitle={International conference on machine learning},
  pages={20503--20521},
  year={2022},
  organization={PMLR}
}

@article{drug_repurposing,
  title={Drug repurposing: progress, challenges and recommendations},
  author={Pushpakom, Sudeep and Iorio, Francesco and Eyers, Patrick A and Escott, K Jane and Hopper, Shirley and Wells, Andrew and Doig, Andrew and Guilliams, Tim and Latimer, Joanna and McNamee, Christine and Norris, Alan and Sanseau, Philippe and Cavalla, David and Pirmohamed, Munir},
  journal={Nature Reviews Drug Discovery},
  volume={18},
  number={1},
  pages={41--58},
  year={2019},
  publisher={Nature Publishing Group}
}

@article{new_drug_old_target,
  title={Development of new drugs for an old target—the penicillin binding proteins},
  author={Zervosen, Astrid and Sauvage, Eric and Fr{\`e}re, Jean-Marie and Charlier, Paulette and Luxen, Andr{\'e}},
  journal={Mol.},
  volume={17},
  number={11},
  pages={12478--12505},
  year={2012},
  publisher={MDPI}
}

@article{new_drug_old_target2,
  title={New drugs, old targets: tweaking the dopamine system to treat psychostimulant use disorders},
  author={Newman, Amy Hauck and Ku, Therese and Jordan, Chloe J and Bonifazi, Alessandro and Xi, Zheng-Xiong},
  journal={Annual review of pharmacology and toxicology},
  volume={61},
  pages={609--628},
  year={2021},
  publisher={Annual Reviews}
}

@article{bindingdb1,
  title={BindingDB: a web-accessible database of experimentally determined protein--ligand binding affinities},
  author={Liu, Tiqing and Lin, Yuhmei and Wen, Xin and Jorissen, Robert N and Gilson, Michael K},
  journal={Nucleic acids research},
  volume={35},
  number={suppl\_1},
  pages={D198--D201},
  year={2007},
  publisher={Oxford University Press}
}

@article{bindingdb2,
   author = {Gilson, Michael K. and Liu, Tiqing and Baitaluk, Michael and Nicola, George and Hwang, Linda and Chong, Jenny},
   title = {BindingDB in 2015: A public database for medicinal chemistry, computational chemistry and systems pharmacology},
   journal = {Nucleic Acids Research},
   volume = {44},
   number = {D1},
   pages = {D1045-D1053},
   ISSN = {0305-1048},
   DOI = {10.1093/nar/gkv1072},
   url = {https://doi.org/10.1093/nar/gkv1072},
   year = {2016},
   type = {Journal Article}
}

@article{EGFR,
  title={Review of epidermal growth factor receptor biology},
  author={Herbst, Roy S},
  journal={International Journal of Radiation Oncology* Biology* Physics},
  volume={59},
  number={2},
  pages={S21--S26},
  year={2004},
  publisher={Elsevier}
}

@article{mpro_inhibitors,
  title={Perspectives on SARS-CoV-2 main protease inhibitors},
  author={Gao, Kaifu and Wang, Rui and Chen, Jiahui and Tepe, Jetze J and Huang, Faqing and Wei, Guo-Wei},
  journal={J. Med. Chem. },
  volume={64},
  number={23},
  pages={16922--16955},
  year={2021},
  publisher={ACS Publications}
}

@article{nedler_mead_opt,
  title={A simplex method for function minimization},
  author={Nelder, John A and Mead, Roger},
  journal={The computer journal},
  volume={7},
  number={4},
  pages={308--313},
  year={1965},
  publisher={The British Computer Society}
}

@article{ECFP,
  title={Extended-connectivity fingerprints},
  author={Rogers, David and Hahn, Mathew},
  journal={J. Chem. Info. Model.},
  volume={50},
  number={5},
  pages={742--754},
  year={2010},
  publisher={ACS Publications}
}

@article{dice_sim,
  title={Measures of the amount of ecologic association between species},
  author={Dice, Lee R},
  journal={Ecology},
  volume={26},
  number={3},
  pages={297--302},
  year={1945},
  publisher={JSTOR}
}

@book{thermodynamics,
  title={Thermodynamics},
  author={Fermi, Enrico},
  year={2012},
  publisher={Courier Corporation}
}

@article{activity_relationship,
  title={Measuring bioactivity: KI, IC50 and EC50},
  author={Mar{\'E}chal, Eric},
  journal={Chemogenomics and Chemical Genetics: A User's Introduction for Biologists, Chemists Informaticians},
  pages={55--65},
  year={2011},
  publisher={Springer}
}

@article{mpro1,
  title={Discovery of S-217622, a noncovalent oral SARS-CoV-2 3CL protease inhibitor clinical candidate for treating COVID-19},
  author={Unoh, Yuto and Uehara, Shota and Nakahara, Kenji and Nobori, Haruaki and Yamatsu, Yukiko and Yamamoto, Shiho and Maruyama, Yuki and Taoda, Yoshiyuki and Kasamatsu, Koji and Suto, Takahiro and Kouki, Kensuke and Nakahashi, Atsufumi and Kawashima, Sho and Sanaki, Takao and Toba, Shinsuke and Uemura, Kentaro and Mizutare, Tohru and Ando, Shigeru and Sasaki, Michihito and Orba, Yasuko and Sawa, Hirofumi and Sato, Akihiko and Sato, Takafumi and Kato, Teruhisa and Tachibana, Yuki},
  journal={J. Med. Chem. },
  volume={65},
  number={9},
  pages={6499--6512},
  year={2022},
  publisher={ACS Publications}
}

@article{mpro2,
  title={Crystal structure of SARS-CoV-2 main protease in complex with non-covalent inhibitor ML188},
  author={Lockbaum, Gordon J and Reyes, Archie C and Lee, Jeong Min and Tilvawala, Ronak and Nalivaika, Ellen A and Ali, Akbar and Kurt Yilmaz, Nese and Thompson, Paul R and Schiffer, Celia A},
  journal={Viruses},
  volume={13},
  number={2},
  pages={174},
  year={2021},
  publisher={MDPI}
}

@article{mpro3,
  title={Anti-SARS-CoV-2 activities in vitro of Shuanghuanglian preparations and bioactive ingredients},
  author={Su, H. X. and Yao, S. and Zhao, W. F. and Li, M. J. and Liu, J. and Shang, W. J. and Xie, H. and Ke, C. Q. and Hu, H. C. and Gao, M. N. and Yu, K. Q. and Liu, H. and Shen, J. S. and Tang, W. and Zhang, L. K. and Xiao, G. F. and Ni, L. and Wang, D. W. and Zuo, J. P. and Jiang, H. L. and Bai, F. and Wu, Y. and Ye, Y. and Xu, Y. C.},
  journal={Acta Pharm. Sin.},
  volume={41},
  number={9},
  pages={1167--1177},
  year={2020},
  publisher={Springer Singapore Singapore}
}

@article{mpro4,
  title={X-ray screening identifies active site and allosteric inhibitors of SARS-CoV-2 MPro},
  author={G{\"u}nther, Sebastian and Reinke, Patrick YA and Fern{\'a}ndez-Garc{\'\i}a, Yaiza and Lieske, Julia and Lane, Thomas J and Ginn, Helen M and Koua, Faisal HM and Ehrt, Christiane and Ewert, Wiebke and Oberthuer, Dominik and Yefanov, Oleksandr and Meier, Susanne and Lorenzen, Kristina and Krichel, Boris and Kopicki, Janine-Denise and Gelisio, Luca and Brehm, Wolfgang and Dunkel, Ilona and Seychell, Brandon and Gieseler, Henry and Norton-Baker, Brenna and Escudero-Pérez, Beatriz and Domaracky, Martin and Saouane, Sofiane and Tolstikova, Alexandra and White, Thomas A. and Hänle, Anna and Groessler, Michael and Fleckenstein, Holger and Trost, Fabian and Galchenkova, Marina and Gevorkov, Yaroslav and Li, Chufeng and Awel, Salah and Peck, Ariana and Barthelmess, Miriam and Schlünzen, Frank and Lourdu Xavier, P. and Werner, Nadine and Andaleeb, Hina and Ullah, Najeeb and Falke, Sven and Srinivasan, Vasundara and França, Bruno Alves and Schwinzer, Martin and Brognaro, Hévila and Rogers, Cromarte and Melo, Diogo and Zaitseva-Kinneberg, Joanna Irina and Knoska, Juraj and Peña-Murillo, Gisel E. and Mashhour, Aida Rahmani and Hennicke, Vincent and Fischer, Pontus and Hakanpää, Johanna and Meyer, Jan and Gribbon, Philip and Ellinger, Bernhard and Kuzikov, Maria and Wolf, Markus and Beccari, Andrea R. and Bourenkov, Gleb and von Stetten, David and Pompidor, Guillaume and Bento, Isabel and Panneerselvam, Saravanan and Karpics, Ivars and Schneider, Thomas R. and Garcia-Alai, Maria Marta and Niebling, Stephan and Günther, Christian and Schmidt, Christina and Schubert, Robin and Han, Huijong and Boger, Juliane and Monteiro, Diana C. F. and Zhang, Linlin and Sun, Xinyuanyuan and Pletzer-Zelgert, Jonathan and Wollenhaupt, Jan and Feiler, Christian G. and Weiss, Manfred S. and Schulz, Eike-Christian and Mehrabi, Pedram and Karničar, Katarina and Usenik, Aleksandra and Loboda, Jure and Tidow, Henning and Chari, Ashwin and Hilgenfeld, Rolf and Uetrecht, Charlotte and Cox, Russell and Zaliani, Andrea and Beck, Tobias and Rarey, Matthias and Günther, Stephan and Turk, Dusan and Hinrichs, Winfried and Chapman, Henry N. and Pearson, Arwen R.},
  journal={Science},
  volume={372},
  number={6542},
  pages={642--646},
  year={2021},
  publisher={American Association for the Advancement of Science}
}

@article{mpro5,
  title={Masitinib is a broad coronavirus 3CL inhibitor that blocks replication of SARS-CoV-2},
  author={Drayman, N. and DeMarco, J. K. and Jones, K. A. and Azizi, S. A. and Froggatt, H. M. and Tan, K. and Maltseva, N. I. and Chen, S. and Nicolaescu, V. and Dvorkin, S. and Furlong, K. and Kathayat, R. S. and Firpo, M. R. and Mastrodomenico, V. and Bruce, E. A. and Schmidt, M. M. and Jedrzejczak, R. and Muñoz-Alía, MÁ and Schuster, B. and Nair, V. and Han, K. Y. and O'Brien, A. and Tomatsidou, A. and Meyer, B. and Vignuzzi, M. and Missiakas, D. and Botten, J. W. and Brooke, C. B. and Lee, H. and Baker, S. C. and Mounce, B. C. and Heaton, N. S. and Severson, W. E. and Palmer, K. E. and Dickinson, B. C. and Joachimiak, A. and Randall, G. and Tay, S.},
  journal={Science},
  volume={373},
  number={6557},
  pages={931--936},
  year={2021},
  publisher={American Association for the Advancement of Science}
}

@article{mpro6,
  title={Expedited approach toward rational design of noncovalent SARS-CoV-2 MPro inhibitors},
  author={Kitamura, N. and Sacco, M. D. and Ma, C. and Hu, Y. and Townsend, J. A. and Meng, X. and Zhang, F. and Zhang, X. and Ba, M. and Szeto, T. and Kukuljac, A. and Marty, M. T. and Schultz, D. and Cherry, S. and Xiang, Y. and Chen, Y. and Wang, J.},
  journal={J. Med. Chem. },
  volume={65},
  number={4},
  pages={2848--2865},
  year={2021},
  publisher={ACS Publications}
}

@article{mpro7,
  title={Potent noncovalent inhibitors of the main protease of SARS-CoV-2 from molecular sculpting of the drug perampanel guided by free energy perturbation calculations},
  author={Zhang, Chun-Hui and Stone, Elizabeth A. and Deshmukh, Maya and Ippolito, Joseph A. and Ghahremanpour, Mohammad M. and Tirado-Rives, Julian and Spasov, Krasimir A. and Zhang, Shuo and Takeo, Yuka and Kudalkar, Shalley N. and Liang, Zhuobin and Isaacs, Farren and Lindenbach, Brett and Miller, Scott J. and Anderson, Karen S. and Jorgensen, William L.},
  journal={ACS central science},
  volume={7},
  number={3},
  pages={467--475},
  year={2021},
  publisher={ACS Publications}
}

@article{mpro8,
  title={Hit expansion of a noncovalent SARS-CoV-2 main protease inhibitor},
  author={Glaser, Jens and Sedova, Ada and Galanie, Stephanie and Kneller, Daniel W. and Davidson, Russell B. and Maradzike, Elvis and Del Galdo, Sara and Labbé, Audrey and Hsu, Darren J. and Agarwal, Rupesh and Bykov, Dmytro and Tharrington, Arnold and Parks, Jerry M. and Smith, Dayle M. A. and Daidone, Isabella and Coates, Leighton and Kovalevsky, Andrey and Smith, Jeremy C.},
  journal={ACS Pharm. Trans. Sci.},
  volume={5},
  number={4},
  pages={255--265},
  year={2022},
  publisher={ACS Publications}
}

@article{mpro9,
  title={Structure-guided design of perampanel-derived pharmacophore targeting SARS-CoV-2 main protease},
  author={Deshmukh, Maya G and Ippolito, Joseph A and Zhang, Chun-Hui and Stone, Elizabeth A and Reilly, Raquel A and Miller, Scott J and Jorgensen, William L and Anderson, Karen S},
  journal={Structure},
  volume={29},
  number={8},
  pages={823--833},
  year={2021},
  publisher={Elsevier}
}

@article{mpro10,
  title={Development of highly potent noncovalent inhibitors of SARS-CoV-2 3CLpro},
  author={Hou, Ningke and Shuai, Lei and Zhang, Lijing and Xie, Xuping and Tang, Kaiming and Zhu, Yunkai and Yu, Yin and Zhang, Wenyi and Tan, Qiaozhu and Zhong, Gongxun and Wen, Zhiyuan and Wang, Chong and He, Xijun and Huo, Hong and Gao, Haishan and Xu, You and Xue, Jing and Peng, Chen and Zou, Jing and Schindewolf, Craig and Menachery, Vineet and Su, Wenji and Yuan, Youlang and Shen, Zuyuan and Zhang, Rong and Yuan, Shuofeng and Yu, Hongtao and Shi, Pei-Yong and Bu, Zhigao and Huang, Jing and Hu, Qi},
  journal={ACS central science},
  volume={9},
  number={2},
  pages={217--227},
  year={2023},
  publisher={ACS Publications}
}

@article{mpro11,
  title={Structural, electronic, and electrostatic determinants for inhibitor binding to subsites S1 and S2 in SARS-CoV-2 main protease},
  author={Kneller, Daniel W. and Li, Hui and Galanie, Stephanie and Phillips, Gwyndalyn and Labbé, Audrey and Weiss, Kevin L. and Zhang, Qiu and Arnould, Mark A. and Clyde, Austin and Ma, Heng and Ramanathan, Arvind and Jonsson, Colleen B. and Head, Martha S. and Coates, Leighton and Louis, John M. and Bonnesen, Peter V. and Kovalevsky, Andrey},
  journal={J. Med. Chem. },
  volume={64},
  number={23},
  pages={17366--17383},
  year={2021},
  publisher={ACS Publications}
}

@article{mpro12,
  title={Discovery and crystallographic studies of trisubstituted piperazine derivatives as non-covalent SARS-CoV-2 MPro inhibitors with high target specificity and low toxicity},
  author={Gao, Shenghua and Sylvester, Katharina and Song, Letian and Claff, Tobias and Jing, Lanlan and Woodson, Molly and Weiße, Renato H. and Cheng, Yusen and Schäkel, Laura and Petry, Marvin and Gütschow, Michael and Schiedel, Anke C. and Sträter, Norbert and Kang, Dongwei and Xu, Shujing and Toth, Karoly and Tavis, John and Tollefson, Ann E. and Müller, Christa E. and Liu, Xinyong and Zhan, Peng},
  journal={J. Med. Chem. },
  volume={65},
  number={19},
  pages={13343--13364},
  year={2022},
  publisher={ACS Publications}
}

@article{mpro13,
  title={Structure-based optimization of ML300-derived, noncovalent inhibitors targeting the severe acute respiratory syndrome coronavirus 3CL protease (SARS-CoV-2 3CLpro)},
  author={Han, Sang Hoon and Goins, Christopher M. and Arya, Tarun and Shin, Woo-Jin and Maw, Joshua and Hooper, Alice and Sonawane, Dhiraj P. and Porter, Matthew R. and Bannister, Breyanne E. and Crouch, Rachel D. and Lindsey, A. Abigail and Lakatos, Gabriella and Martinez, Steven R. and Alvarado, Joseph and Akers, Wendell S. and Wang, Nancy S. and Jung, Jae U. and Macdonald, Jonathan D. and Stauffer, Shaun R.},
  journal={J. Med. Chem. },
  volume={65},
  number={4},
  pages={2880--2904},
  year={2021},
  publisher={ACS Publications}
}

@article{egfr1,
  title={Structures of lung cancer-derived EGFR mutants and inhibitor complexes: mechanism of activation and insights into differential inhibitor sensitivity},
  author={Yun, Cai-Hong and Boggon, Titus J and Li, Yiqun and Woo, Michele S and Greulich, Heidi and Meyerson, Matthew and Eck, Michael J},
  journal={Cancer cell},
  volume={11},
  number={3},
  pages={217--227},
  year={2007},
  publisher={Elsevier}
}

@article{egfr2,
  title={Structural insights into drug development strategy targeting EGFR T790M/C797S},
  author={Su-Jie Zhu and Peng Zhao and Jiao Yang and Rui Ma and Xiao-E Yan and Sheng-yong Yang and Jing-Wen Yang and Cai‐Hong Yun},
  journal={Oncotarget},
  year={2018},
  volume={9},
  pages={13652 - 13665},
  url={https://api.semanticscholar.org/CorpusID:4090455}
}

@article{egfr3,
  title={Structural basis of mutant-selectivity and drug-resistance related to CO-1686},
  author={Yan, Xiao-E and Zhu, Su-Jie and Liang, Ling and Zhao, Peng and Choi, Hwan Geun and Yun, Cai-Hong},
  journal={Oncotarget},
  volume={8},
  number={32},
  pages={53508},
  year={2017},
  publisher={Impact Journals, LLC}
}

@article{egfr4,
  title={Protein kinase inhibitor design by targeting the Asp-Phe-Gly (DFG) motif: the role of the DFG motif in the design of epidermal growth factor receptor inhibitors},
  author={Peng, Yi-Hui and Shiao, Hui-Yi and Tu, Chih-Hsiang and Liu, Pang-Min and Hsu, John Tsu-An and Amancha, Prashanth Kumar and Wu, Jian-Sung and Coumar, Mohane Selvaraj and Chen, Chun-Hwa and Wang, Sing-Yi and Lin, Wen-Hsing and Sun, Hsu-Yi and Chao, Yu-Sheng and Lyu, Ping-Chiang and Hsieh, Hsing-Pang and Wu, Su-Ying},
  journal={J. Med. Chem. },
  volume={56},
  number={10},
  pages={3889--3903},
  year={2013},
  publisher={ACS Publications}
}

@article{egfr5,
  title={4-Aminoindazolyl-dihydrofuro [3, 4-d] pyrimidines as non-covalent inhibitors of mutant epidermal growth factor receptor tyrosine kinase},
  author={Hanan, E. J. and Baumgardner, M. and Bryan, M. C. and Chen, Y. and Eigenbrot, C. and Fan, P. and Gu, X. H. and La, H. and Malek, S. and Purkey, H. E. and Schaefer, G. and Schmidt, S. and Sideris, S. and Yen, I. and Yu, C. and Heffron, T. P.},
  journal={Bioorg. Med. Chem. Lett.},
  volume={26},
  number={2},
  pages={534--539},
  year={2016},
  publisher={Elsevier}
}

@article{egfr6,
  title={Structure-based approach for discovery of pyrrolo [3, 2-d] pyrimidine-based EGFR T790M/L858R mutant inhibitors},
  author={Sogabe, Satoshi and Kawakita, Youichi and Igaki, Shigeru and Iwata, Hidehisa and Miki, Hiroshi and Cary, Douglas R and Takagi, Terufumi and Takagi, Shinji and Ohta, Yoshikazu and Ishikawa, Tomoyasu},
  journal={ACS Med. Chem. Lett.},
  volume={4},
  number={2},
  pages={201--205},
  year={2013},
  publisher={ACS Publications}
}

@article{egfr7,
  title={Structure of the epidermal growth factor receptor kinase domain alone and in complex with a 4-anilinoquinazoline inhibitor},
  author={Stamos, Jennifer and Sliwkowski, Mark X and Eigenbrot, Charles},
  journal={J. Bio. Chem.},
  volume={277},
  number={48},
  pages={46265--46272},
  year={2002},
  publisher={ASBMB}
}

@article{egfr8,
  title={Pyridones as highly selective, noncovalent inhibitors of T790M double mutants of EGFR},
  author={Bryan, Marian C. and Burdick, Daniel J. and Chan, Bryan K. and Chen, Yuan and Clausen, Saundra and Dotson, Jennafer and Eigenbrot, Charles and Elliott, Richard and Hanan, Emily J. and Heald, Robert and Jackson, Philip and La, Hank and Lainchbury, Michael and Malek, Shiva and Mann, Sam E. and Purkey, Hans E. and Schaefer, Gabriele and Schmidt, Stephen and Seward, Eileen and Sideris, Steve and Wang, Shumei and Yen, Ivana and Yu, Christine and Heffron, Timothy P.},
  journal={ACS Med. Chem. Lett.},
  volume={7},
  number={1},
  pages={100--104},
  year={2016},
  publisher={ACS Publications}
}

@article{egfr9,
  title={Protein kinase inhibitor design by targeting the Asp-Phe-Gly (DFG) motif: the role of the DFG motif in the design of epidermal growth factor receptor inhibitors},
  author={Peng, Yi-Hui and Shiao, Hui-Yi and Tu, Chih-Hsiang and Liu, Pang-Min and Hsu, John Tsu-An and Amancha, Prashanth Kumar and Wu, Jian-Sung and Coumar, Mohane Selvaraj and Chen, Chun-Hwa and Wang, Sing-Yi and Lin, Wen-Hsing and Sun, Hsu-Yi and Chao, Yu-Sheng and Lyu, Ping-Chiang and Hsieh, Hsing-Pang and Wu, Su-Ying},
  journal={J. Med. Chem. },
  volume={56},
  number={10},
  pages={3889--3903},
  year={2013},
  publisher={ACS Publications}
}

@article{egfr10,
  title={Structural analysis of the mechanism of inhibition and allosteric activation of the kinase domain of HER2 protein},
  author={Aertgeerts, K. and Skene, R. and Yano, J. and Sang, B. C. and Zou, H. and Snell, G. and Jennings, A. and Iwamoto, K. and Habuka, N. and Hirokawa, A. and Ishikawa, T. and Tanaka, T. and Miki, H. and Ohta, Y. and Sogabe, S.},
  journal={J. Bio. Chem.},
  volume={286},
  number={21},
  pages={18756--18765},
  year={2011},
  publisher={ASBMB}
}

@article{egfr11,
  title={Discovery of n-((3 r, 4 r)-4-fluoro-1-(6-((3-methoxy-1-methyl-1 h-pyrazol-4-yl) amino)-9-methyl-9 h-purin-2-yl) pyrrolidine-3-yl) acrylamide (pf-06747775) through structure-based drug design: A high affinity irreversible inhibitor targeting oncogenic egfr mutants with selectivity over wild-type egfr},
  author={Planken, Simon and Behenna, Douglas C. and Nair, Sajiv K. and Johnson, Theodore O. and Nagata, Asako and Almaden, Chau and Bailey, Simon and Ballard, T. Eric and Bernier, Louise and Cheng, Hengmiao and Cho-Schultz, Sujin and Dalvie, Deepak and Deal, Judith G. and Dinh, Dac M. and Edwards, Martin P. and Ferre, Rose Ann and Gajiwala, Ketan S. and Hemkens, Michelle and Kania, Robert S. and Kath, John C. and Matthews, Jean and Murray, Brion W. and Niessen, Sherry and Orr, Suvi T. M. and Pairish, Mason and Sach, Neal W. and Shen, Hong and Shi, Manli and Solowiej, James and Tran, Khanh and Tseng, Elaine and Vicini, Paolo and Wang, Yuli and Weinrich, Scott L. and Zhou, Ru and Zientek, Michael and Liu, Longqing and Luo, Yiqin and Xin, Shuibo and Zhang, Chengyi and Lafontaine, Jennifer},
  journal={J. Med. Chem. },
  volume={60},
  number={7},
  pages={3002--3019},
  year={2017},
  publisher={ACS Publications}
}

@article{egfr12,
  title={The T790M mutation in EGFR kinase causes drug resistance by increasing the affinity for ATP},
  author={Yun, Cai-Hong and Mengwasser, Kristen E and Toms, Angela V and Woo, Michele S and Greulich, Heidi and Wong, Kwok-Kin and Meyerson, Matthew and Eck, Michael J},
  journal={Proc. Natl. Acad. Sci.},
  volume={105},
  number={6},
  pages={2070--2075},
  year={2008},
  publisher={National Acad Sciences}
}

@article{egfr13,
  title={Discovery of novel 4-amino-6-arylaminopyrimidine-5-carbaldehyde oximes as dual inhibitors of EGFR and ErbB-2 protein tyrosine kinases},
  author={Xu, Guozhang and Searle, Lily Lee and Hughes, Terry V. and Beck, Amanda K. and Connolly, Peter J. and Abad, Marta C. and Neeper, Michael P. and Struble, Geoffrey T. and Springer, Barry A. and Emanuel, Stuart L. and Gruninger, Robert H. and Pandey, Niranjan and Adams, Mary and Moreno-Mazza, Sandra and Fuentes-Pesquera, Angel R. and Middleton, Steven A. and Greenberger, Lee M.},
  journal={Bioorg. Med. Chem. Lett.},
  volume={18},
  number={12},
  pages={3495--3499},
  year={2008},
  publisher={Elsevier}
}

@article{egfr14,
  title={Binding mode of breakthrough inhibitor AZD9291 to EGF receptor revealed},
  author={Yosaatmadja, Yuliana and Silva, Shevan and Dickson, James M and Patterson, Adam V and Smaill, Jeff B and Flanagan, Jack U and McKeage, Mark J and Squire, Christopher J},
  journal={J. Struct. Bio.},
  volume={192},
  number={3},
  pages={539--544},
  year={2015},
  publisher={Elsevier}
}

@article{egfr15,
  title={A unique structure for epidermal growth factor receptor bound to GW572016 (Lapatinib) relationships among protein conformation, inhibitor off-rate, and receptor activity in tumor cells},
  author={Wood, E. R. and Truesdale, A. T. and McDonald, O. B. and Yuan, D. and Hassell, A. and Dickerson, S. H. and Ellis, B. and Pennisi, C. and Horne, E. and Lackey, K. and Alligood, K. J. and Rusnak, D. W. and Gilmer, T. M. and Shewchuk, L.},
  journal={Cancer research},
  volume={64},
  number={18},
  pages={6652--6659},
  year={2004},
  publisher={AACR}
}

@article{pdb_3v3m,
  title={Discovery, synthesis, and structure-based optimization of a series of N-(tert-butyl)-2-(N-arylamido)-2-(pyridin-3-yl) acetamides (ML188) as potent noncovalent small molecule inhibitors of the severe acute respiratory syndrome coronavirus (SARS-CoV) 3CL protease},
  author={Jacobs, Jon and Grum-Tokars, Valerie and Zhou, Ya and Turlington, Mark and Saldanha, S. Adrian and Chase, Peter and Eggler, Aimee and Dawson, Eric S. and Baez-Santos, Yahira M. and Tomar, Sakshi and Mielech, Anna M. and Baker, Susan C. and Lindsley, Craig W. and Hodder, Peter and Mesecar, Andrew and Stauffer, Shaun R.},
  journal={J. Med. Chem. },
  volume={56},
  number={2},
  pages={534--546},
  year={2013},
  publisher={ACS Publications}
}

@article{RFScorev3,
  title={Improving AutoDock Vina using random forest: the growing accuracy of binding affinity prediction by the effective exploitation of larger data sets},
  author={Li, Hongjian and Leung, Kwong-Sak and Wong, Man-Hon and Ballester, Pedro J},
  journal={Molecular informatics},
  volume={34},
  number={2-3},
  pages={115--126},
  year={2015},
  publisher={Wiley Online Library}
}

@article{fingerprint,
  title={Proteo-chemometrics interaction fingerprints of protein--ligand complexes predict binding affinity},
  author={Wang, Debby D and Xie, Haoran and Yan, Hong},
  journal={Bioinformatics},
  volume={37},
  number={17},
  pages={2570--2579},
  year={2021},
  publisher={Oxford University Press}
}

@article{generalization,
  title={Assessment of the generalization abilities of machine-learning scoring functions for structure-based virtual screening},
  author={Zhu, Hui and Yang, Jincai and Huang, Niu},
  journal={Journal of Chemical Information and Modeling},
  volume={62},
  number={22},
  pages={5485--5502},
  year={2022},
  publisher={ACS Publications}
}

@incollection{pytorch,
  title     = {PyTorch: An Imperative Style, High-Performance Deep Learning Library},
  author    = {Paszke, Adam and Gross, Sam and Massa, Francisco and Lerer, Adam and Bradbury, James and Chanan, Gregory and Killeen, Trevor and Lin, Zeming and Gimelshein, Natalia and Antiga, Luca and Desmaison, Alban and Kopf, Andreas and Yang, Edward and DeVito, Zachary and Raison, Martin and Tejani, Alykhan and Chilamkurthy, Sasank and Steiner, Benoit and Fang, Lu and Bai, Junjie and Chintala, Soumith},
  booktitle = {Advances in Neural Information Processing Systems 32},
  pages     = {8024--8035},
  year      = {2019},
  publisher = {Curran Associates, Inc.},
  url       = {http://papers.neurips.cc/paper/9015-pytorch-an-imperative-style-high-performance-deep-learning-library.pdf}
}

@article{scikit-learn,
  title={Scikit-learn: Machine Learning in {P}ython},
  author={Pedregosa, F. and Varoquaux, G. and Gramfort, A. and Michel, V.
          and Thirion, B. and Grisel, O. and Blondel, M. and Prettenhofer, P.
          and Weiss, R. and Dubourg, V. and Vanderplas, J. and Passos, A. and
          Cournapeau, D. and Brucher, M. and Perrot, M. and Duchesnay, E.},
  journal={Journal of Machine Learning Research},
  volume={12},
  pages={2825--2830},
  year={2011}
}

@article{vina_new_weights,
  title={Improving ligand-ranking of AutoDock Vina by changing the empirical parameters},
  author={Pham, T. N. H. and Nguyen, T. H. and Tam, N. M. and T, Y. Vu and Pham, N. T. and Huy, N. T. and Mai, B. K. and Tung, N. T. and Pham, M. Q. and V, V. Vu and Ngo, S. T.},
  journal={Journal of Computational Chemistry},
  volume={43},
  number={3},
  pages={160--169},
  year={2022},
  publisher={Wiley Online Library}
}

@misc{CanonizedRMSD,
      title={Canonized then Minimized RMSD for Three-Dimensional Structures}, 
      author={Jie Li and Qian Chen and Jingwei Weng and Jianming Wu and Xin Xu},
      year={2024},
      eprint={2405.00339},
      archivePrefix={arXiv},
      primaryClass={physics.chem-ph}
}

@article{kldiv,
  title={On information and sufficiency},
  author={Kullback, Solomon and Leibler, Richard A},
  journal={The annals of mathematical statistics},
  volume={22},
  number={1},
  pages={79--86},
  year={1951},
  publisher={JSTOR}
}

@article{PERT,
  title={PERT-perfect random tree ensembles},
  author={Cutler, Adele and Zhao, Guohua},
  journal={Computing Science and Statistics},
  volume={33},
  number={4},
  pages={90--4},
  year={2001}
}

@article{overfitting_or_perfect_fitting,
  title={Overfitting or perfect fitting? risk bounds for classification and regression rules that interpolate},
  author={Belkin, Mikhail and Hsu, Daniel J and Mitra, Partha},
  journal={Advances in neural information processing systems},
  volume={31},
  year={2018}
}

@article{gorantla2023proteins,
  title={From Proteins to Ligands: Decoding Deep Learning Methods for Binding Affinity Prediction},
  author={Gorantla, Rohan and Kubincova, Alzbeta and Wei{\ss}e, Andrea Y and Mey, Antonia SJS},
  journal={Journal of Chemical Information and Modeling},
  year={2023},
  volume = {64},
   number = {7},
   pages = {2496-2507},
  publisher={ACS Publications}
}

@article{volkov2022frustration,
  title={On the frustration to predict binding affinities from protein--ligand structures with deep neural networks},
  author={Volkov, Mikhail and Turk, Joseph-Andr{\'e} and Drizard, Nicolas and Martin, Nicolas and Hoffmann, Brice and Gaston-Math{\'e}, Yann and Rognan, Didier},
  journal={Journal of medicinal chemistry},
  volume={65},
  number={11},
  pages={7946--7958},
  year={2022},
  publisher={ACS Publications}
}

@article{GIGN,
  title={Geometric interaction graph neural network for predicting protein--ligand binding affinities from 3d structures (gign)},
  author={Yang, Ziduo and Zhong, Weihe and Lv, Qiujie and Dong, Tiejun and Yu-Chian Chen, Calvin},
  journal={The journal of physical chemistry letters},
  volume={14},
  number={8},
  pages={2020--2033},
  year={2023},
  publisher={ACS Publications}
}

@article{biopython,
  title={Biopython: freely available Python tools for computational molecular biology and bioinformatics},
  author={Cock, Peter J. A. and Antao, Tiago and Chang, Jeffrey T. and Chapman, Brad A. and Cox, Cymon J. and Dalke, Andrew and Friedberg, Iddo and Hamelryck, Thomas and Kauff, Frank and Wilczynski, Bartek and de Hoon, Michiel J. L.},
  journal={Bioinformatics},
  volume={25},
  number={11},
  pages={1422},
  year={2009},
  publisher={Oxford University Press}
}

@article{similarity_scores,
  title={Why is Tanimoto index an appropriate choice for fingerprint-based similarity calculations?},
  author={Bajusz, D{\'a}vid and R{\'a}cz, Anita and H{\'e}berger, K{\'a}roly},
  journal={Journal of cheminformatics},
  volume={7},
  pages={1--13},
  year={2015},
  publisher={Springer}
}

\end{document}


\section{Reformulation of the PDBBind Dataset} 

\subsection{Cleaning the PDBBind dataset}
The majority of protein-ligand affinity prediction models are designed for non-peptide drug-like small molecules. However, a careful investigation of the protein-ligand complexes in PDBBind dataset elucidated that all the ligands are not necessarily drug like. Including these data in the training of a ML model might make it harder for the model to learn key features related to the binding of drug-like small molecules to protein targets. Furthermore, some proteins and ligands in the PDBBind dataset contain uncommon elements, such as Hg, Cu and Sr in proteins or Co, Se or V in ligands. A small portion of the structures in PDBBind dataset also contained steric clashes, or ligands far away from any peptide chain (for example, PDB ID: 2R1W). These were  excluded from the dataset due to the low quality of the complex structures, and the PDB IDs containing steric clashes (minimum heavy atom distance $<$1.75 Å) are also provided below. It is difficult for ML models to learn protein-ligand interactions when poor data and uncommon elements are present in the PDBBind dataset. Therefore, we have defined a "cleaner" version of PDBBind which only contains all data with ligand QED values larger than 0.2, protein and ligand elements that have at least 1‰ (more than 19 occurrences in the dataset), and minimum heavy-atom distance between protein and ligand fall within 1.75 and 4 Å. We call this subset of the data Clean Level 1 (CL1). 

The binding affinity in terms of $\Delta G$ is directly related to the dissociation coefficient $K_d$ or $K_i$ through the formula $\Delta G=-RT \ln(K)$.\cite{thermodynamics} However, a large portion of the data in PDBBind is reported in terms of $\mathrm{IC}_{50}$, which cannot be easily translated to $\Delta G$s due to its dependence on other experimental conditions and inhibition mechanisms.\cite{activity_relationship} The $\mathrm{IC}_{50}$ values for the same protein-ligand complex can vary up to one order of magnitude in different assays. In addition, some data in PDBBind were not reported as exact values. Therefore, a second clean level (CL2) was defined on top of CL1, that additionally requires the target values are converted from records with “$K_d=\mathrm{xxx}$” or “$K_i=\mathrm{xxx}$”, to ensure the reliability of experimental binding free energy data. 

The total amount of data in PDBBind v2020 filtered with CL1 and CL2 are 14324 and 7985, respectively. The CL1 level dataset was used for retraining models described later because it contains the most amount of data. The CL2 level dataset was used for validation and testing because the binding affinity data are more reliable at this level. 

\subsection{Description for processing the PDBBind dataset}  All PDB files were downloaded from RCSB to recover the original headers, and the categories of the proteins were defined according to whether the keywords occurred in the header. The header information was extracted using the biopython package version 1.81.\cite{biopython} The categories we considered include transport proteins, hydrolases, transferases, transcription proteins, lyases, oxidoreductases, isomerases, ligases, membrane proteins, viral proteins, chaperone proteins, and metalloproteins. A protein file with none of these keywords occurring in the header were categorized into a generic "other" category. Protein sequences were extracted directly from the PDBBind dataset files using the SEQRES records, and sequences for proteins with multiple chains were concatenated using a column (:) symbol. Additionally, the SMILES strings for each of the ligands were extracted using the rdkit package (2023.3.2 version)\cite{rdkit} from either the .mol2 or .sdf file provided in the original PDBBind dataset. The .mol2 files were used with higher priority due to their generally better description of bond orders in the ligands compared with the .sdf files in the PDBBind dataset. The latter were used instead when rdkit failed to read in .mol2 files.

\subsection{Similarity calculation of protein and ligands in the PDBBind dataset}
Similarities for each ligand to any other ligand in the whole PDBBind dataset were calculated based on the Morgan fingerprints of the ligands using 1024 bits\cite{ECFP}. According to a study that compares multiple similarity score metrics, the Tanimoto index, Dice index, Cosine coefficient and Soergel distance were giving similar results on molecule similarity calculations and all worked pretty well.\cite{similarity_scores} In our work the Dice similarity is reported for the ligand pairs according to the following\cite{dice_sim}:
\begin{equation}
    sim=\frac{2|A\cup B|}{|A|+|B|}
\end{equation} where $|A\cup B|$ counts the number of bits set to ON in the fingerprints of both ligands A and B, and $|X|$ counts the number of bits set to ON in an single ligand X. A radius of 2 was first used when calculating Morgan fingerprints of the ligands; if the similarity between two ligands was calculated to be 1, it is recalculated using Morgan fingerprints of the ligands with radius of 4 to allow more careful validation of ligand identity, and the radius is extended to 10 if a radius of 4 still result in a similarity of 1. If the similarity was still 1 with the extended radius for calculating the fingerprint, the canonical SMILES string of the two molecules were compared and any discrepancy in the canonical SMILES will enforce the similarity to 0.99. These steps were designed to ensure ligands with similarity of 1 are strictly identical.

Similarities for proteins were calculated based on the aligned sequences of the proteins. Considering that it is unlikely two proteins belonging to different functional categories are similar, the sequence similarities were calculated only for proteins in the same category (i.e. transport proteins, hydrolases, etc.), and any two proteins that belong to different categories were defined to have similarity of 0. Within each category, every pair of protein sequences were aligned using the Needleman-Wunsch alignment algorithm\cite{needleman_wunsch}, and the similarity was calculated as the number of aligned residues divided by the total length of the aligned sequence.

\subsection{Data Resplitting of protein and ligands in the PDBBind dataset}
Data splitting was first done inside each protein category, and an iterative process was employed to separate the dataset step by step using the following algorithm: In the first iteration, 5 complexes were randomly selected as seed data in the test set, and all data in the same protein category that have protein sequence similarity greater than 0.9 or ligand similarity greater than 0.99 were added to the test set as well. If protein-ligand complex data that have protein sequence similarity greater than 0.5 or ligand similarity greater than 0.99 were found, they too were added to the validation set. Any data newly added to the test set in this iteration will become seed data in the next iteration, and the iteration continues until no new data are added to the test set. Next, a similar process was applied to the validation data, adding all remaining data that have protein sequence similarity greater than 0.5 or ligand similarity greater than 0.99 into the validation set. The remaining data is then defined as the training set.

After data splitting by protein category, the protein-ligand complexes for training, validation, and testing were combined to define the splitting for the whole dataset. However, after combination the ligand similarity might still exceed 0.99 for data from different categories. Therefore, any data in the combined training set that has ligand similarity greater than 0.99 to any other data in the validation or test set were discarded altogether. The resulting number of data in the training, validation, and test set were 11513, 2422 and 4860, respectively, after this cleaning step.

% \subsection{Compilation of the BDB2020+ dataset}
% Since a complex may contain multiple ligands, but only one matches the record in BindingDB, we selected the ligand in the PDB that has the best structural match with the SMILEs provided in BindingDB, and ensuring that the number of heavy atoms is exactly the same. We then used rdkit to reassign bond orders to the extracted ligands using the BindingDB SMILES as reference. This step was necessary because bond orders are usually not present in a PDB file and are typically inferred from local atomic geometries, which sometimes result in unreasonable bonding structures; thus the bond order reassignment step ensures the rationality of the processed structures. Additionally, any chain that is within 5Å of the ligand was compared with the interacting chain sequence in BindingDB record. A reliable match was only made when the consecutive aligned residues are exactly the same. The reason to keep strict alignment criterion is that if the protein contains mutations, the binding affinity might change significantly, in which case the BindingDB record will not represent the true binding affinity for the complex structure in the PDB and is not usable for the benchmark. After discarding all unmatched data, we obtained 130 data records, out of which 115 contains accurate binding affinity data, and defines the BDB2020+ test dataset.

\subsection{AutoDock Vina Retraining}
There are six empirical parameters in the AutoDock Vina scoring function (gauss1, gauss2, repulsion, hydrophobic, hydrogen, and rotation penalty) \cite{autodock_vina,AD-vina2021}. For each molecule in the LP-PDBBind training set, the 6 individual terms of AutoDock Vina were calculated using the vina binary (v1.2.3) by setting weight of one term to one and the rest of the weight to 0. The weights of these six terms were optimized by minimizing the mean absolute error between the weighted sum of six terms and binding free energy through the Nelder-Mead optimization algorithm\cite{nedler_mead_opt}. The final retrained AutoDock Vina parameters are provided in Supplementary Table 1 and all further evaluations of retrained models are done with these modified weights.

\subsection{RF-Score Retraining}
The RF-Score (RF) uses Random Forest model to predicts binding affinities with counts of a particular protein-ligand atom type pair interacting within a certain distance range based on 3D structures\cite{rfscore}. We here followed the RF-Score-v1 approach, where nine common elemental atom types (C, N, O, F, P, S, Cl, Br, I) for both the protein and the ligand were considered and neighboring contacts between a protein-ligand atom pair were defined within 12 Å. Only atom pairs with non-zero occurrence over the whole training dataset were considered, which result in 36 features in total. For fairness of comparison, we trained a RF model on the PDBbind2007 refined set, denoted as the original model, and on the LP-PDBBind training data, as the retrained model, using the same script. 
%For the original model, we reported a pearson R of 0.974 on the training data and a pearson R of 0.689 on the PDBbind 2007 core set as test data.

\subsection{IGN Retraining}
InteractionGraphNet (IGN)\cite{IGN} utilizes two independent graph convolution modules to sequentially learn the intramolecular and intermolecular interactions from the 3D structures of protein-ligand complexes to predict binding affinities \cite{IGN}.  Protein pockets were extracted using Chimera (v1.17.3) by selecting residues that are within 10 Å of any atom from the ligand, and then the protein and ligand structures were saved as RDKit objects. Due to molecule generation errors in RDKit \cite{rdkit}, which is required for featurizing the 3D structures into graph representations, only 6378 complexes from the LP-PDBBind training set were used. The code for retraining the IGN model (https://github.com/THGLab/LP-PDBBind/tree/master/model\_retraining/IGN) was adapted from the original training scripts using the same feature size and layer numbers as the original published model with only the training data modified. The exact hyperparameters for retraining are provided below:

\subsection{DeepDTA Retraining}
DeepDTA is a purely data-driven approach which does not rely on physical interactions or 3D structural information. Instead, it uses 1D convolutions on the string representations of the proteins and ligands to make predictions.\cite{deepDTA}  Since the method did not originally train on the PDBBind dataset, we retrained the model with the original PDBBind general and refined sets using a PyTorch implementation of the original code: https://github.com/THGLab/LP-PDBBind/tree/master/model\_retraining/deepdta.   More specifically, we did a 90-10 train-validation split on the data for training and then tested its performance on the core set. Different protein and ligand kernel sizes were used as suggested by the original work and the best-performed parameters with the protein kernel size of 12, ligand kernel size of 4 and channel size of 32 were chosen to serve as the original model in our performance comparison tables. The retrained model with LP-PDBBind was also generated following the same scheme.

Since the 1d convolution requires a fixed size of the protein sequence and ligands, during all training processes, proteins with sequence lengths longer than 2000 and ligands with isomeric SMILES lengths longer than 200 were discarded, resulting in the final number of training data, validation data and test data to be 7338, 956 and 2162, respectively. Also, since the ligand encoding was based on the training set only, all ligands with unseen tokens from their isomeric SMILES strings in the test set and real-world examples are discarded, resulting in a loss of 15 data points. It is worth noting that these losses of data happen to both the original model and the retrained model, so it won't affect our conclusions in this work.

\section{Covalent Binders}
The identifiers of the 893 covalent binders in PDBBind v2020 (count: 893):
\\
\noindent 3HJ0, 2XLC, 6Q35, 4BCB, 5GWZ, 6P8X, 4DKT, 3B1T, 3BG8, 5E0G, 1EAS, 5WEJ, 1Z6F, 1BIO, 6B1H, 3BJM, 5I24, 3ZCZ, 4JJE, 6N4T, 4AXM, 2I72, 4QVM, 1KDW, 3HD3, 6HVS, 4R3B, 1P06, 5TG4, 5DGJ, 5DPA, 2OC1, 1P04, 5FAO, 4QWU, 5ZWH, 4WKS, 2O7V, 4GS6, 5C1Y, 1A3E, 4NO8, 1QX1, 2G5T, 1KE3, 3SN8, 4QWL, 3ZMH, 5T66, 3KJF, 5D9P, 5DP5, 3D4F, 1MXO, 1L6Y, 5WDL, 1LHE, 1PI4, 4BCC, 3LJ7, 1RHM, 5WAG, 6DA4, 2QL5, 6BL1, 5OM9, 6B1J, 5GWY, 1O45, 5OYD, 1QWU, 4OON, 1KDS, 6CHA, 6B1F, 4QWS, 6HHH, 6BKX, 5H6V, 3KW9, 1ONG, 4QVL, 5X02, 5J7S, 4AN1, 5FAT, 1P10, 1RHQ, 2H5D, 6RNI, 4DCD, 6DGE, 4WKV, 4AMZ, 5FA7, 6RNE, 4QZ2, 1QCP, 3P8E, 2AUX, 1I8J, 1K2I, 2FDA, 1O4K, 3NS7, 6B0Y, 1VSN, 5TYN, 1AU0, 1RWX, 1NC6, 5CYI, 4JG8, 4D8E, 3LLE, 1F1J, 3PA8, 6M9F, 3SNB, 4LEN, 4OOL, 4MNV, 1BGO, 1ZPC, 6B1Y, 3UR9, 6CN8, 4KQO, 1AHT, 2FS9, 6MKQ, 4GK7, 3NZI, 1QJ6, 4BS5, 4E3L, 4J70, 6HVU, 4PIQ, 1P01, 1F9E, 5INH, 6FV1, 3S1Y, 1RHJ, 4QW0, 5W14, 4MAO, 5DP6, 5HL9, 5ACB, 2MLM, 4O7D, 6ND3, 2WIJ, 4S2I, 6AFJ, 5KRE, 3TJM, 5TYL, 1GZG, 2G5P, 2WJ2, 2ZZ6, 4QVV, 3BM8, 2XNI, 3H0E, 2QNZ, 5SYS, 1Q6K, 4WX6, 5MJB, 4PKB, 6HVW, 1RTL, 3BH3, 4QW5, 5DP4, 1NY0, 3OYP, 1E34, 1U9W, 1YMX, 3SVV, 6SKD, 2WJ1, 4XBB, 5V2Q, 3W2P, 5VQY, 9LPR, 6DUD, 3LCE, 1GA9, 6HHI, 3RDH, 4Z16, 1YT7, 2XU5, 4ZRO, 5FOO, 1NMS, 2WOQ, 1FSW, 6HV7, 6NVG, 2NQG, 6UN3, 1QJ1, 5TTU, 3MXR, 6Q5B, 6CZU, 4YQV, 5JH6, 3SNA, 4WYY, 4IMZ, 5EEC, 6EYZ, 4F49, 5T6F, 2Y59, 4E3I, 3IKA, 6I0X, 1AYV, 4QZ3, 2XDM, 3KWB, 4LV3, 5L6O, 1RHR, 3ZMI, 6RN6, 8LPR, 4QWX, 4WKU, 1AU2, 1TMB, 3MBZ, 4PL5, 6FFS, 4XJR, 2LPR, 6FDU, 5TH7, 5HLB, 5MAJ, 2HWP, 1UOD, 1A46, 6NVI, 1GFW, 6QG7, 4WX4, 4J5P, 6QHR, 6FDQ, 3LPR, 5FAQ, 1MPL, 3VB5, 1FSY, 6MHB, 4X68, 3V4J, 6IUO, 6HVR, 4QPS, 6DB4, 4QBB, 1ORW, 2OZ2, 3EX6, 5WFJ, 6M8Y, 4X6J, 5VQV, 1NQC, 4EST, 6D3G, 6QHO, 1GMY, 6AFE, 2R4B, 2ALV, 5V2P, 4QW4, 4IMQ, 4XBD, 6HHG, 5J9Z, 2Y2N, 4YV8, 1INC, 5E7R, 5NE1, 3PR0, 6AFA, 3FMQ, 3O6T, 4OB2, 3FMR, 2WZX, 3W2Q, 6B95, 1O4I, 6RNU, 3QN7, 6S1S, 3KQA, 5VQX, 5U4G, 1MY8, 1ZPB, 4WKT, 3SZB, 6UN1, 6QW8, 4X69, 3ZEB, 4JG7, 6IB2, 3LOK, 6HUB, 7LPR, 6B1X, 1NYY, 5J87, 6IC6, 4WSK, 5SWH, 3V4X, 2FTD, 5YU9, 3VB7, 3O87, 4QVN, 4JK6, 5DG6, 6PNO, 6IYV, 5J8I, 3K84, 6NVJ, 5HG8, 2Y2I, 5VQE, 3ZVW, 2A4Q, 6HTD, 6BIC, 4WZ5, 1YLV, 3KJN, 5C91, 4JR0, 3UFA, 6S9W, 1IEM, 3A73, 1AD8, 6AFI, 5W13, 5J8X, 5WKL, 5NUD, 4E3N, 5X5G, 4YQU, 3BM6, 4QWI, 1H8Y, 2WGI, 5LCJ, 4WEF, 3QSD, 1TBZ, 2QLB, 1GGD, 2WIK, 4QVW, 1PAU, 6LPR, 4WM9, 4QVP, 6B41, 2XU1, 5TDI, 1ZOM, 3LOX, 2XDW, 6NVL, 2QQ7, 5DP8, 1YLY, 6M8W, 5XHR, 3SZ9, 4YAS, 1O4D, 3EWU, 2QKY, 5KYK, 6HVA, 1P02, 1ERO, 3GJS, 1W31, 2OP9, 3OPR, 1O43, 6HTR, 5GMP, 4VGC, 3OPP, 5WAD, 4GD6, 4NO1, 5LC0, 6O8I, 4WSJ, 4OLC, 1U9X, 6JPJ, 4QZ1, 3ZOT, 6K1S, 2XLN, 2XOW, 6G9F, 3VB6, 2EEP, 4QWR, 4MVN, 5LPR, 5TG2, 3FNM, 1MNS, 4MZO, 6DI1, 6MHM, 2AUZ, 1B0F, 2AJD, 3D62, 2Q9N, 1C3B, 5TYJ, 6RJP, 6PNN, 2Q3Z, 5J9Y, 4QXJ, 1P03, 6MZW, 5TTS, 3QKV, 3BLT, 5UG9, 3ZMJ, 2JAL, 3O1G, 4NNN, 1QHR, 4Q2K, 4PNC, 3KWZ, 4QQC, 6M9C, 3SV8, 1W12, 5V88, 6E5G, 2QLJ, 4JJ7, 2Y2K, 5ORL, 1A5G, 4I9O, 4NNW, 3ZVT, 2QLQ, 6M9D, 5TOZ, 5TYK, 6OVZ, 4QWJ, 2QLF, 3GPJ, 2Y4A, 2AJL, 3V6R, 4LV2, 2GVF, 1EXW, 4MNW, 4YHF, 6HHJ, 1BMQ, 4UUQ, 2GBG, 4WX7, 5ZWF, 4XUZ, 5HLD, 6AFL, 5D6F, 1IEW, 4TWY, 2O9A, 3C9E, 2HOB, 5YOF, 5L6P, 5TG6, 1VGC, 3O86, 1RHU, 4GHT, 4QZ4, 3G3M, 6CQZ, 1O4A, 5FQ9, 5EST, 5A3H, 1DOJ, 4E3J, 3K83, 5E0H, 4PIS, 5GTY, 3S22, 3S3Q, 5TEH, 1GBT, 1ERQ, 1NL6, 2XU4, 2H5I, 1A09, 4PL3, 2WIG, 1LHD, 4QW7, 7GCH, 2ZU4, 4QZX, 1W10, 4NO9, 5D11, 6F6R, 1LLB, 6DQB, 5TYP, 6E5B, 4E3K, 1LHG, 5LWN, 1LHF, 1RXP, 4MZS, 2ZU5, 4M1J, 6GOP, 6AFG, 1RWW, 4YRT, 4QWF, 1H1B, 5GNK, 4I7D, 3SND, 4PID, 5E0J, 1NJU, 1KE0, 5WAC, 2OC0, 1MS6, 2UZJ, 2AJB, 5JK3, 5TTV, 3HHA, 1NO9, 1NPZ, 5X79, 6PGP, 4INT, 2QL9, 5D6E, 2QCN, 6S9X, 3G3D, 6MHC, 3BLS, 1B5G, 4INH, 2OP3, 3MXS, 3VGC, 1RE1, 4XCU, 5UG8, 5VQZ, 2FM2, 6NVH, 2F9U, 3S3R, 1L6S, 2GBF, 4HCV, 4PL4, 3W2T, 6AFC, 4I7C, 6HUQ, 1G37, 6AFH, 2YJB, 5DP7, 5U4F, 4M8T, 5DP9, 4QVY, 3KRD, 4JMX, 2XU3, 4WBG, 4NO6, 2F9V, 2BDL, 6RN9, 2Y2J, 2YJC, 4QZ0, 4CCD, 1W14, 6N9P, 4QWG, 4PJI, 5GSO, 4HRD, 6MNY, 6G8N, 2V6N, 4MNX, 3BLU, 6HUC, 6G7F, 2XK1, 6OIM, 5WKM, 5FAS, 4E3M, 4D8I, 5TG1, 6HTC, 5V4Q, 4WMC, 6ALZ, 3HWN, 4MBF, 2HWO, 4QZ5, 1P05, 3MKF, 5C1X, 6CQT, 6B1O, 4YEC, 6QFT, 1EAT, 4HRC, 5WAE, 1IAU, 2ZU3, 6IYW, 3RJM, 1U9V, 1AUJ, 2QL7, 4HNP, 2Y2P, 6IC5, 3E90, 1YM1, 6FV2, 5W12, 1TYN, 6BL2, 4QKX, 3ZIM, 1QTN, 4AMY, 2NQI, 3OF8, 3EX3, 2H65, 5NE3, 1LHC, 1EKB, 1SNK, 5AHJ, 1MEM, 4KIO, 6ARY, 3OVX, 4INR, 5ZWE, 1NLJ, 1RHK, 4HCU, 2FXR, 4AMX, 6BIB, 4U0X, 6QW9, 1YK7, 2I03, 3LXS, 3SNC, 1ONH, 4QW1, 1TU6, 3B1U, 6B1W, 6HGY, 5UGC, 2Z3Z, 6MHD, 4MZ4, 4MNY, 5VND, 4JK5, 2QAF, 3OJ8, 6GCH, 1MWT, 6P8Y, 3KJQ, 4YRS, 6HUV, 2RCX, 2H5J, 5LVX, 4QWK, 5F02, 6D8E, 5WAF, 5EE8, 4QW3, 4NK3, 4JG6, 5G0Q, 2G63, 2Y55, 6B0V, 5MAE, 4R02, 5T6G, 5LCK, 6HUU, 5HG9, 6AFD, 6ERT, 4R6V, 4INU, 1A61, 6QG4, 5NGF, 4FZG, 5FAP, 4QQ5, 2WAP, 6GCR, 6BID, 5TYO, 4HBP, 2OBO, 6RTN, 5TG5, 4QZ7, 3EYD, 1PI5, 5MQY, 3SV7, 6HVT, 6GXY, 2FJ0, 4JJ8, 6RN7, 2XE4, 6QWB, 3VB4, 2A4G, 2Q9M, 4BSQ, 4AMW, 5ZA2, 3U1I, 3FKV, 6H0U, 4QW6, 3T9T, 1O41, 2FS8, 1E37, 3I4A, 6HV5, 4YQM, 4LV1, 6B1E, 2Y2H, 3BWK, 2YJ9, 1O4E, 5I23, 2OC7, 1AYU, 6FFN, 1QJ7, 5XYZ, 6AFF, 5TG7, 3BAR, 3I06, 5GWA, 5J5D, 4E3O, 6HVV, 4QZ6, 6PGO, 4WZ4, 2XZC, 1HBJ, 5HG7, 6G9S, 6AX1, 2LP8, 6GZY, 2YJ8, 4Q1S, 6AF9, 6QW7, 4L0L, 3SJO, 1AWF, 1F7B, 1TLO, 1QFS, 6IB0, 4FZC, 5NPB, 5NWZ, 6QWA, 1YMS, 4QZW, 2WZZ, 3IUT, 6SKB, 6J6M, 5C20, 1AWH, 2VGC, 1NKM, 6HV4, 6G8M, 6BQ0, 2R6N, 5HG5, 3N4C, 3K7F, 6PNM, 3SJI, 4IVK, 4QVQ, 4DMY, 6P8Z, 4RSP, 5CLS, 3O88, 2YJ2, 3V6S, 4X0U, 4AN0, 6HTP, 2XCN, 3SV6

\section{Structures Containing Steric Clashes in PDBBind}
Here are the PDBIDs for structures in PDBBind v2020 with minimum distance between heavy atoms in the ligand and the protein less than 1.75Å (count: 1303):
\noindent 3N9S, 4U1B, 3ZHX, 1EW8, 2FZK, 2Q8Z, 5GOF, 1EX8, 1OLX, 3A6T, 1KM3, 4K55, 5EFA, 3K1J, 4L51, 3NXQ, 6D1H, 1FZQ, 1JLR, 6O5T, 1I9N, 4U6C, 2ZZ2, 1X8R, 4U71, 2RK8, 2V8W, 6HH3, 4FL2, 1RNM, 3ISS, 3DJV, 1PVN, 4U6W, 1B38, 1MAI, 1V48, 2C94, 1LCP, 3B7I, 1V11, 4EU0, 4R3W, 6CZB, 4NXV, 6CZC, 6D1G, 1NKI, 3B7R, 4K5P, 2V2H, 1W4Q, 1MRS, 1AFL, 2ISW, 4AXD, 3HU3, 3EVD, 1M83, 4DEL, 3RV4, 3IAE, 6D1B, 4UOH, 3B3C, 1WUQ, 3WGG, 4U0W, 3RBU, 3AHO, 6JDI, 1G98, 3A1E, 4K3N, 4QJX, 2OI2, 1JVU, 1B55, 2W8W, 6IHT, 2RCN, 2R75, 3D2E, 3PCG, 2GSU, 4NCN, 4KAX, 2C92, 1TLP, 4QRH, 2YAY, 1TKB, 3S8L, 1YQ7, 1G53, 1KSN, 4DCS, 1JYQ, 1QK4, 4P5D, 1FT7, 6D9S, 4PVY, 6OF5, 2QTN, 5D6J, 2AMT, 4L50, 4IDN, 6D1J, 1WUR, 1RNT, 1SLN, 1BXR, 3G1V, 1PKX, 4ZCS, 5CBM, 1LVU, 6JON, 3I73, 3EBI, 3EXH, 2TMN, 5HVA, 4U69, 3OVE, 2WF5, 4JNE, 4FL1, 2E91, 1M5W, 5FSX, 2XBP, 4U73, 5HWU, 2WE3, 6FAA, 6D15, 1RP7, 6D3Q, 5NW7, 1M7Y, 1YFZ, 1U0G, 2PY4, 4LPS, 1YDD, 5LTN, 3DJO, 3K5X, 2QTA, 3BXF, 5TMP, 2RIO, 1PFU, 2PU1, 1B57, 3A1C, 3D7K, 1S5Z, 1M0Q, 1A4R, 6J72, 3UPK, 4U70, 1V16, 6MNV, 3JUK, 4U6Z, 1ATR, 6FHK, 3OZG, 1OLS, 3EGT, 3AHN, 3OV1, 3AAS, 6DD0, 3Q71, 3EXE, 3FZN, 6D1A, 1D2E, 2E92, 3N1C, 1MUE, 5EVZ, 4ETZ, 4JYC, 6D18, 2PQC, 3C56, 6HH5, 5C2O, 3FZY, 1HYO, 5BV3, 2NSL, 6H77, 3I9G, 5UV2, 1HI4, 3DJP, 2R59, 4DY6, 1EW9, 2VT3, 2FXV, 5YHE, 5LDP, 1Q54, 2WZF, 3FUC, 2W5G, 2PTZ, 5F2R, 5HJQ, 1X8T, 2X97, 4OR6, 1VSO, 3MKE, 3HJ0, 2XLC, 6Q35, 4BCB, 5KHG, 3LOO, 3L0K, 5TIG, 6JN4, 2Y4M, 6EJI, 2PJ4, 4QIR, 3BG8, 1QF5, 6G15, 1EAS, 1Z6F, 1BIO, 6B1H, 3BJM, 5I24, 3ZCZ, 5YBI, 3HU2, 6N4T, 2I72, 4QVM, 1KDW, 3POA, 1TMM, 4EFG, 6HVS, 3GGC, 2VF6, 4R3B, 1SRE, 1P06, 5O3R, 5TG4, 5JNL, 1NU8, 2OC1, 3ATV, 1P04, 5FAO, 5IZM, 4QWU, 2CFG, 6N97, 5EOU, 4WKS, 4OCP, 5C1Y, 3UIG, 2Y1G, 4NO8, 1ONP, 2G5T, 1KE3, 4QWL, 3ZMH, 5T66, 3KR5, 5D9P, 1DB4, 1WC6, 3D4F, 4F3H, 1MXO, 1L6Y, 1Y8P, 1LHE, 2PJA, 1PI4, 4ZQT, 4BCC, 3LJ7, 6CZD, 5JEK, 5WAG, 3TDH, 6DA4, 3UED, 3WNS, 5OM9, 6B1J, 3EWZ, 1O45, 3D9P, 5OYD, 1QWU, 1H79, 4DXJ, 4OON, 1KDS, 6B1F, 4UMJ, 4QWS, 1TKC, 6BKX, 5H6V, 1ONG, 5ZDG, 4KXM, 6I5J, 4P4F, 4QVL, 5X02, 5J7S, 5J7J, 4AN1, 4DN0, 4P4S, 5FAT, 1T29, 1P10, 1R0X, 2H5D, 5B5P, 1QJB, 4WKV, 4AMZ, 5FA7, 6FEL, 6RNE, 1Y8O, 4QZ2, 1QCP, 3GQO, 2MC1, 3ZP9, 2AUX, 2PJC, 1I8J, 3E0P, 1K2I, 6G01, 2FDA, 1O4K, 6AEC, 5W38, 3QXC, 6B22, 6B0Y, 1VSN, 6FA5, 4YX9, 5TYN, 4KP4, 6NNG, 5CYI, 4BXN, 4G8L, 3PA8, 6FAC, 6IPL, 6M9F, 4K5N, 4LEN, 1AO0, 4OOL, 4MNV, 4E1E, 1ZPC, 1KWR, 6B1Y, 3FYZ, 5L44, 4YB7, 6R4V, 1CSI, 1H7A, 2Z7H, 1OTH, 4KQO, 5F1C, 4JJF, 4EEJ, 1AHT, 2FS9, 6MKQ, 4GK7, 3NZI, 1QJ6, 4E3L, 2J4K, 4J70, 6HVU, 1P01, 5INH, 1H07, 3S1Y, 4QW0, 5W14, 5DP6, 5HL9, 4WYZ, 5ACB, 1TPZ, 4O7D, 3X1K, 2WIJ, 4S2I, 5ICV, 5KRE, 4R17, 3TJM, 2JBV, 5TYL, 4QLU, 2QPJ, 3TL0, 2G5P, 2WJ2, 2ZZ6, 4QVV, 2XNI, 2QNZ, 6AGP, 4PKB, 6HVW, 1LF8, 1RTL, 3BH3, 4QW5, 1EBG, 1NY0, 1E34, 1U9W, 2IOA, 1YMX, 4K8O, 6SKD, 2Z7I, 2WJ1, 1TL7, 1X07, 2Y1D, 9LPR, 1GA9, 1MN9, 3WDC, 6AK6, 4QLV, 6R4S, 6FMP, 3RDH, 3DPC, 5FOO, 2WOQ, 1FSW, 2PJ5, 6HV7, 5AB0, 3RYW, 6UN3, 1QJ1, 5TTU, 3L0N, 6AEH, 3MXR, 6Q5B, 1CSS, 6CZU, 4R18, 1G05, 5JH6, 6OM4, 4JG0, 4WYY, 4IMZ, 5EEC, 6EYZ, 3HU1, 2VHQ, 5NPF, 2Y59, 4E3I, 6PK7, 5UV1, 5Z68, 4K5M, 6IPM, 4QZ3, 2XDM, 4LV3, 2C1N, 3AAV, 4FUT, 4FR3, 3ZMI, 1PPW, 8LPR, 5BRN, 1NZV, 4QWX, 5F29, 4WKU, 3EXF, 6N54, 4GKC, 4RAB, 3MBZ, 4PL5, 4JJG, 2XAQ, 2P0X, 4XJR, 6IPH, 2LPR, 6FDU, 4U6E, 5HLB, 3HWX, 1UOD, 4DEM, 1P6D, 1AKQ, 1LV8, 1AMN, 6QG7, 4J5P, 6FDQ, 5NPR, 3FHE, 3LPR, 5FAQ, 1MPL, 5L6H, 3VB5, 1FSY, 4X68, 3V4J, 4QLQ, 6HVR, 1O8B, 6NZG, 4HWT, 6DB4, 1ML1, 1AKV, 5JAZ, 1ORW, 2OZ2, 3EX6, 5WFJ, 6M8Y, 4X6J, 1B39, 1NQC, 5UQ9, 4EST, 5W10, 4JN4, 6D3G, 1EIX, 1RYH, 6N94, 4QW4, 2A4R, 5YR5, 1KF0, 1D7X, 2QTR, 5EY8, 5JMP, 5J9Z, 2Y2N, 4YV8, 1INC, 3T2C, 3LXO, 6MLF, 1E8H, 5NE1, 3PR0, 1JBD, 3FMQ, 4OB2, 3FMR, 2WZX, 1O4I, 6RNU, 6S1S, 5U4G, 1MY8, 1YHM, 2C9D, 3D67, 5YR6, 1ZPB, 4WKT, 6UN1, 6QW8, 4X69, 3ZEB, 6IB2, 6HUB, 7LPR, 3ZKF, 4UV9, 5K6S, 2Z4W, 6B1X, 1NYY, 5IZL, 4WSK, 3VB7, 3K41, 3O87, 5BSK, 4QVN, 3L6F, 3CST, 5OD5, 3K84, 1VJC, 2Y2I, 5VQE, 3ZVW, 2A4Q, 4WUP, 6HTD, 5F5B, 1C7F, 4LOJ, 4WZ5, 3SHV, 1YLV, 2QM7, 1YON, 1VJD, 1U2R, 3ZY2, 3E16, 6Q30, 2Y4L, 1IEM, 1AD8, 6EUV, 5W13, 2QCD, 4I9S, 4GA3, 5AGJ, 5J8X, 4L2X, 4BG6, 5NUD, 3TDZ, 5KHD, 3AU6, 1GGN, 4E3N, 5X5G, 4EHM, 3GZN, 5NHZ, 3BM6, 4QWI, 1H8Y, 2WQP, 2WGI, 6J7L, 1RU2, 4NZM, 1TBZ, 3AXK, 5YKP, 4LIL, 4HE9, 1RGK, 1GGD, 5NZ2, 2WIK, 6RMM, 4QVW, 6LPR, 4WM9, 4QVP, 5GSW, 2PU0, 1ZOM, 6PXC, 5FB0, 1Y19, 3LOX, 1T3T, 4CLP, 2XDW, 1YLY, 6M8W, 5XHR, 4YAS, 1O4D, 1I9L, 3EWU, 2QKY, 2ORK, 5KYK, 6ISD, 6HVA, 6DD1, 1P02, 1ERO, 6GFM, 1AKU, 1W31, 3OPR, 1O43, 4RHU, 2PJ0, 6HTR, 6AR2, 3MLE, 4VGC, 6HYS, 3OPP, 3LLM, 4U0G, 5WAD, 1I9Q, 4EXZ, 4Y67, 4GD6, 4H38, 4NO1, 5LC0, 6E5S, 4FL3, 6AK5, 4WSJ, 1U9X, 4QZ1, 5NYZ, 3WDZ, 3ZOT, 3ZMQ, 6K1S, 2XLN, 2XOW, 3WNR, 6G9F, 3V7D, 3KC0, 3VB6, 4QWR, 3NCQ, 4MVN, 5LPR, 4A0J, 3RG2, 3Q7P, 1QJA, 3FNM, 3IQV, 3IUC, 1MNS, 3RBM, 4OAZ, 2AUZ, 6IA7, 4G5Y, 6G2N, 1B0F, 2AJD, 3D62, 2Q9N, 1C3B, 6GIU, 5TYJ, 6RJP, 1BWN, 2Q3Z, 1I9M, 1NJT, 5J9Y, 4U0U, 2L0I, 4KW6, 4QXJ, 1NZL, 2Y8L, 1P03, 4ZCW, 3EX2, 3QKV, 3BLT, 4UVA, 3ZMJ, 3NBA, 2JAL, 4NNN, 5NPS, 1QHR, 6C5J, 4Q2K, 4W5J, 4EDE, 4LQY, 4QQC, 2E95, 4GS9, 4I9R, 6M9C, 4P5E, 3SV8, 1W12, 2V0C, 2HQU, 4OEL, 6AK4, 2E98, 1O4F, 5J7P, 2Y2K, 3QT7, 2O1V, 4NNW, 3ZS1, 3ATU, 3ZVT, 2QLQ, 1RU1, 6M9D, 5ZDC, 5TYK, 6B4H, 6OVZ, 2HA0, 4QWJ, 3GPJ, 2Y4A, 3OB2, 2H5A, 2AJL, 1BJR, 3Q72, 4LV2, 2CMC, 2GVF, 5HH4, 4UUQ, 2GBG, 5W8V, 4WX7, 2ZA3, 5ZWF, 4XUZ, 1AKR, 5HLD, 3CF1, 1MF4, 5D6F, 4WN0, 2PJ6, 1IEW, 6BS5, 2HOB, 4DWG, 5YOF, 5TG6, 1VGC, 5DOH, 3O86, 1X6U, 4QZ4, 3G3M, 6CQZ, 1O4A, 5FQ9, 5EST, 4PRY, 5A3H, 1DOJ, 4E3J, 3K83, 4AY6, 1GVK, 6IPJ, 3S22, 5TEH, 3WYJ, 1GBT, 6PGX, 4W4S, 1ERQ, 2E93, 3G4F, 4N1Z, 1A09, 4PL3, 2WIG, 3EHW, 1LHD, 1AZL, 4QW7, 7GCH, 5MXQ, 2Z52, 2O4H, 3BH8, 4QZX, 1W10, 4NO9, 5D11, 1LLB, 5TYP, 5IM3, 1CSH, 6E5B, 4E3K, 1LHG, 2A29, 1NFS, 1Y98, 3U8D, 1LHF, 1RXP, 4M1J, 6GOP, 5HYX, 4QWF, 5YR4, 1H1B, 6QXS, 3A1S, 4PID, 4GV8, 1NJU, 2QCG, 1KE0, 5WAC, 2OC0, 6IPI, 6EGS, 3Q2G, 2AJB, 5W19, 5JK3, 1NO9, 5MOC, 6E7M, 5MI3, 5SVK, 5X79, 4INT, 4RXC, 3FCK, 1LRT, 5H63, 5D6E, 2QCN, 3G3D, 6AI9, 3V0P, 3BLS, 1H9L, 2OP3, 3MXS, 3VGC, 6NPF, 5L6J, 2FM2, 2F9U, 1VKJ, 2Q80, 4TMF, 1L6S, 5EWZ, 2GBF, 4PL4, 1RYF, 3W2T, 1XFV, 6HUQ, 5WRS, 1G37, 1CSR, 4JBS, 1MAU, 3QX8, 5DP7, 4M8T, 4QVY, 4K5O, 3KRD, 4JMX, 5AEL, 4RAQ, 4WBG, 6N9T, 5OLK, 2PJ2, 2K0X, 4NO6, 1FPY, 2F9V, 2BDL, 6RN9, 4K19, 2Y2J, 4QZ0, 3E4A, 4CCD, 1W14, 6N9P, 6RML, 4U82, 4QWG, 4PJI, 5GSO, 1GX4, 5M9D, 4HRD, 6G8N, 3BB1, 4QLS, 2E99, 3BLU, 6HUC, 6G7F, 3IOI, 2XK1, 5EY9, 4GZ3, 3UX0, 3UO9, 5FAS, 4E3M, 6HTC, 5V4Q, 4WMC, 3PJT, 4MBF, 4QZ5, 1P05, 3MKF, 5C1X, 6CQT, 1NYM, 6B1O, 6QFT, 1GX0, 3CPH, 5HH6, 1EAT, 1U0H, 4HRC, 4EUV, 5WAE, 1IAU, 2H1H, 1OXG, 3RJM, 1AUJ, 5EXX, 4HNP, 2Y2P, 6IC5, 3E90, 6H78, 3D9L, 1YM1, 5G1P, 4RAC, 5W12, 1TYN, 6BL2, 2ITK, 2G83, 3FV7, 3ZIM, 1QTN, 4AMY, 5HH5, 3EX3, 5NE3, 1LHC, 5A3R, 1EKB, 5AHJ, 5E2V, 3BUO, 6ARY, 6MMC, 4INR, 2FXR, 4AMX, 4U0X, 6QW9, 2P59, 2I03, 1BSK, 3D9M, 1ONH, 4QW1, 4K5L, 1I3Z, 6B1W, 2QCF, 3BU6, 2P1C, 1FT4, 5VND, 4YM4, 2QAF, 4DWB, 2PLL, 3RSB, 3LDW, 16PK, 3OJ8, 6GCH, 1MWT, 2QCM, 4V11, 6HUV, 2RCX, 5LVX, 4QWK, 2JG8, 5WAF, 5EE8, 4QW3, 4NK3, 1ELS, 2IT4, 5G0Q, 2WVA, 3B9S, 2G63, 2Y55, 6B0V, 5MAE, 4R02, 6IPK, 6HUU, 1YYY, 6ERT, 4R6V, 4INU, 1G9R, 5OSY, 4FZG, 2JBJ, 5FAP, 4QQ5, 2WAP, 5TYO, 5X9H, 4HBP, 2OBO, 6RTN, 5TG5, 1I9O, 4QZ7, 3EYD, 1PI5, 3SV7, 6HVT, 2FJ0, 5E2N, 2XE4, 6QWB, 4KXL, 3VB4, 2A4G, 2Q9M, 4BSQ, 5ZA2, 2WIC, 6G6X, 3U1I, 4JFX, 5KNU, 3FKV, 5JBI, 4QW6, 1O41, 2FS8, 1E37, 6CSE, 5LVF, 4OSF, 6HV5, 4LV1, 6B1E, 2Y2H, 3KSL, 1O4E, 5I23, 2OC7, 4RUU, 1QJ7, 5XYZ, 5KNR, 5K7H, 1SZ0, 6BUU, 5TG7, 2AD5, 3BAR, 4AY5, 5GWA, 4E3O, 6NNR, 6HVV, 3FZC, 4QZ6, 1M9N, 4WZ4, 2XZC, 1HBJ, 6G9S, 6AX1, 5EJE, 1TKA, 6GZY, 2L75, 3FSJ, 4OEM, 4Q1S, 3NCR, 5L6I, 6MU1, 6QW7, 4L0L, 5NI0, 4QLT, 1AWF, 1F7B, 6AEJ, 1QFS, 6IB0, 4FZC, 6OTT, 5NPB, 6QWA, 2F89, 1BA8, 2PSX, 4RAO, 3ZJC, 1YMS, 4QZW, 2WZZ, 2Y8Q, 6CJJ, 4GE5, 6SKB, 1AWH, 3HXE, 5JC1, 2VGC, 1SYO, 1NKM, 6HV4, 6G8M, 6BQ0, 3UEC, 5ZDE, 1JRS, 3K7F, 4IVK, 3WNT, 3BGM, 4QVQ, 2HR6, 5CLS, 3O88, 4P4T, 4X0U, 6E3P, 2ONB, 4AN0, 6HTP, 2XCN, 3SV6

\clearpage

\section{Supporting Tables}

\begin{table}[H]
    \centering
    \caption{Number of train, validation and test data in each protein category after creating the new split of PDBBind}
    \begin{tabular}{lccc}
        \hline\hline

\textbf {Protein type } & \textbf { Train } & \textbf { Validation } & \textbf { Test } \\ \hline
hydrolase  & 4038 & 318 & 1377 \\
transferase  & 2226 & 1228 & 1837 \\
other  & 3050 & 405 & 134 \\
transcription  & 642 & 52 & 298 \\
lyase  & 331 & 71 & 465 \\
transport  & 503 & 35 & 83 \\
oxidoreductase  & 342 & 52 & 182 \\
ligase  & 342 & 38 & 90 \\
isomerase  & 199 & 70 & 64 \\
chaperone  & 50 & 66 & 184 \\
membrane  & 157 & 56 & 71 \\
viral  & 196 & 18 & 53 \\
metal containing  & 85 & 13 & 22 \\ 
discarded (too similar) & -648 & - & - \\
        \hline
    \end{tabular}
\label{tab:meanstd}
\end{table}

\begin{table}[ht]
    \centering
    \begin{tabular}{l r}
        \hline
        \hline
        \textbf{Term} & \textbf{Weight} \\
        \hline
        gauss1 & 0.003372 \\
        gauss2 & -0.008098 \\
        repulsion & 0.014212 \\
        hydrophobic & -0.008361 \\
        hydrogen & -0.227928 \\
        rot & 0.05846 \\
        \hline
        \hline
    \end{tabular}
    \caption{Retrained AutoDock Vina Weights}
    \label{stable:vina_weights}
    \label{tab:weights}
\end{table}

\begin{table}[ht]
    \centering
    \begin{tabular}{l r}
        \hline
        \hline
        \textbf{Term} & \textbf{Weight} \\
        \hline
        epochs & 500 \\
        batch\_size & 128 \\
        graph\_feat\_size & 128 \\
        num\_layers & 2 \\
        outdim\_g3 & 128 \\
        d\_FC\_layer & 128 \\
        repetitions & 3 \\
        lr & 0.001 \\
        l2 & 0.00001 \\
        dropout & 0.2 \\
        \hline
        \hline
    \end{tabular}
    \caption{Retrained IGN hyperparameters}
    \label{tab:ign_hyperparams}
\end{table}

\clearpage

\section{Supporting Figures}

\begin{figure}[H]
\begin{center}
\includegraphics[width=0.98\textwidth]{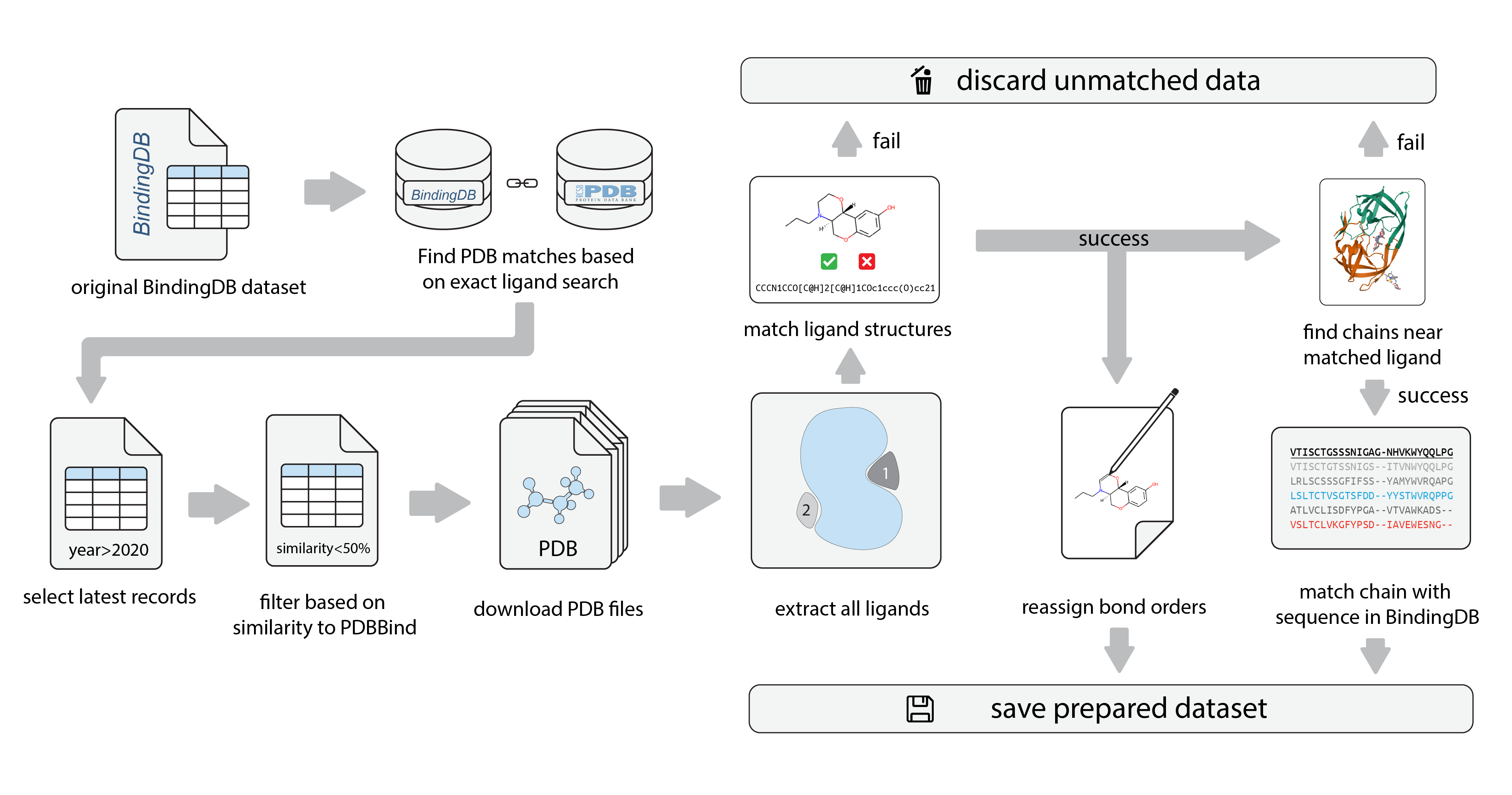}
\end{center}
\caption{\textit{Flowchart of the BDB2020+ curation process.} Starting with the original BindingDB dataset, the records with potential matched structures in the RCSB database were identified by searching the InChi keys that occur in BindingDB to find all PDB records that contain the same ligand, and then further filtering based on the PDBID of the target chain that was also recorded in the BindingDB dataset. Release dates of the PDB structures were extracted from the RCSB PDB records and only records after year 2020 were selected for further processing. Additionally, we employed the same similarity criterion as what was used in developing LP-PDBBind, and removed all data that has sequence similarity greater than 50\% or ligand similarity greater than 99\%. The filtered PDB files were downloaded from the RCSB database and small molecule ligands were extracted from the PDB files. The original BindingDB dataset\cite{bindingdb1,bindingdb2} was matched with PDB database records by exact ligand matches. The matched records were further filtered based on release date and similarity with the PDBBind dataset. All ligands were extracted from the downloaded PDB files, and the ligand matching the SMILES in the BindingDB dataset was selected for further process. Protein chains in the vicinity of the ligand were also extracted. Both the matched ligand and protein chains were further processed to ensure identity with the record in PDBBind. Any discrepancy will result in a mismatch of data and will be discarded.}
\label{fig:bindingDB_curaion}
\end{figure}

\begin{figure}[H]
\begin{center}
\includegraphics[width=\textwidth]{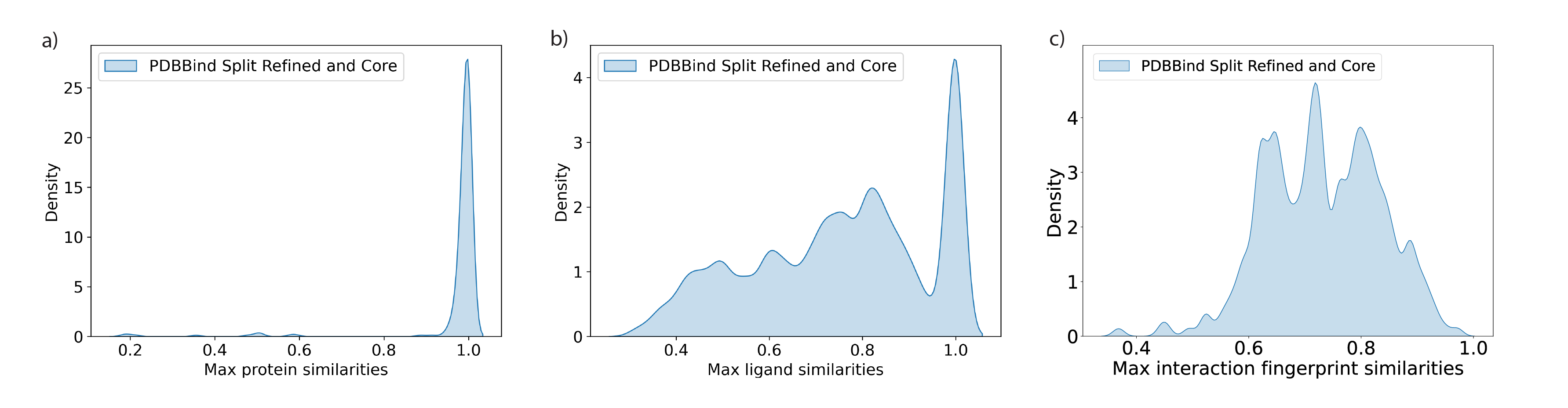}
\end{center}
\vspace{-7mm}
\caption{\textit{Data similarity between refined set and core set in the PDBBind original split.} Comparison of maximum protein similarities
(a), ligand similarities (b) and interaction fingerprint similarities (c) between refined set and core set under original split of PDBBind.}
\label{fig:pdbbind-lig-dist}
\end{figure}

\begin{figure}[H]
\begin{center}
\includegraphics[width=0.98\textwidth]{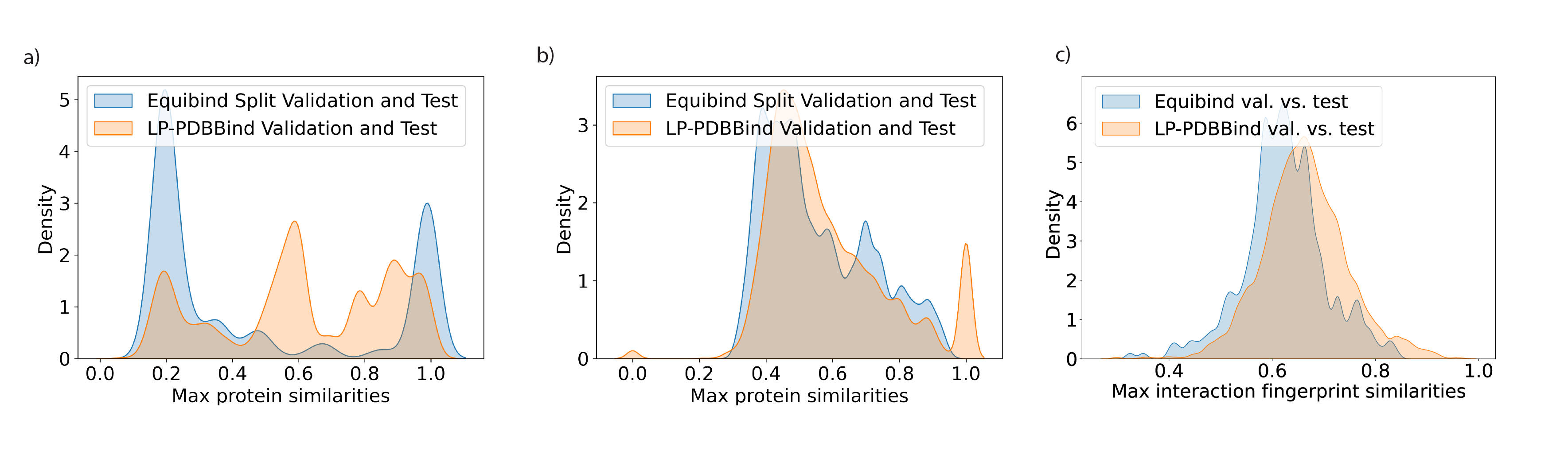}
\end{center}
\vspace{-7mm}
\caption{\textit{Data similarity between validation set and test set in different splittings of PDBBind.} Comparison of maximum protein similarities
(a), ligand similarities (b) and interaction fingerprint similarities (c) between validation set and test set for Equibind and LP-PDBBind.}
\label{fig:pdbbind-lig-dist}
\end{figure}

\begin{figure}[H]
\begin{center}
\includegraphics[width=1.05\textwidth]{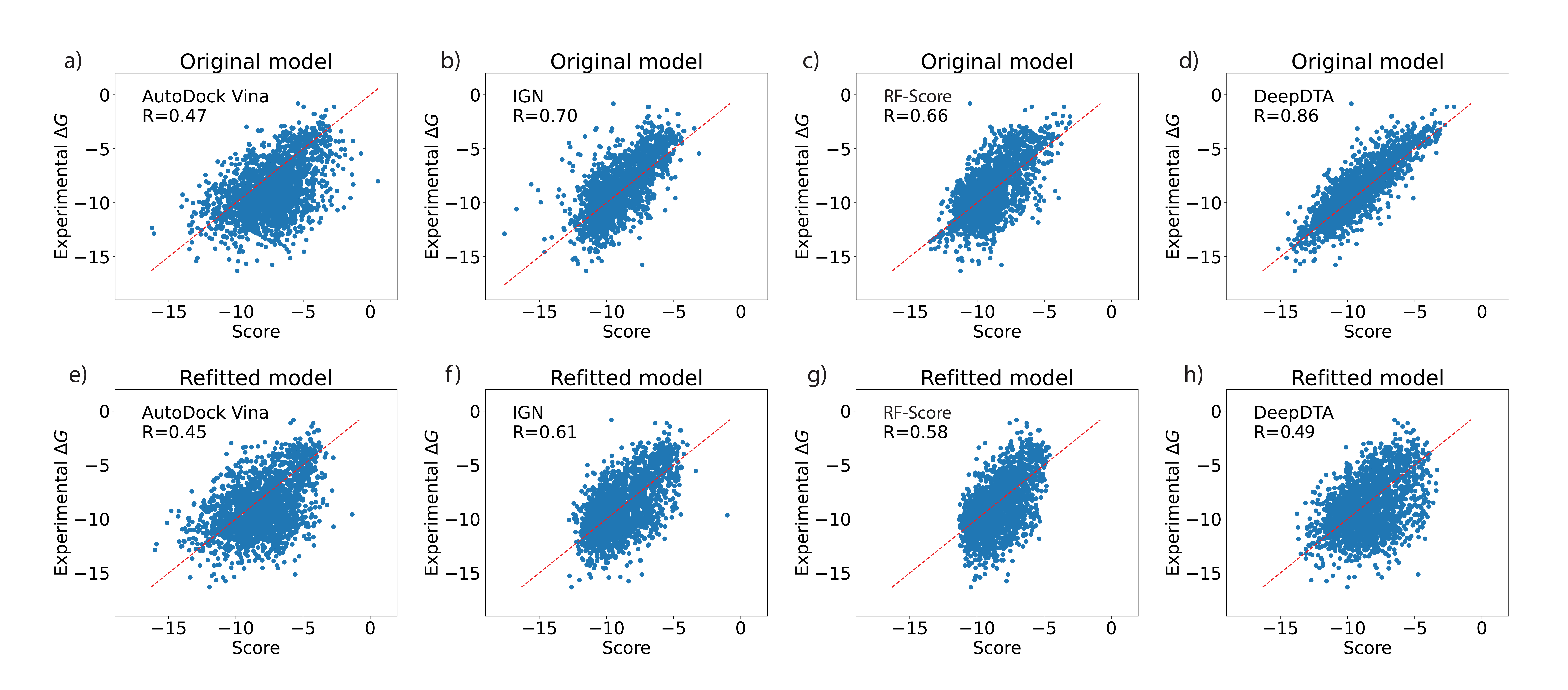}
\end{center}
\vspace{-7mm}
\caption{\textit{Scatter plots for predicted scores and experimental binding affinities before and after model retraining for different models evaluated on UCBSplit test dataset.} Scatter plots for original models: (a) AutoDock vina, (b) IGN, (c) RF-Score, and (d) DeepDTA; Scatter plots for models retrained with UCBSplit: (e) AutoDock vina, (f) IGN, (g) RF-Score, and (h) DeepDTA.}
\label{fig:test_scatters}
\end{figure}

\begin{figure}[H]
\begin{center}
\includegraphics[width=1.05\textwidth]{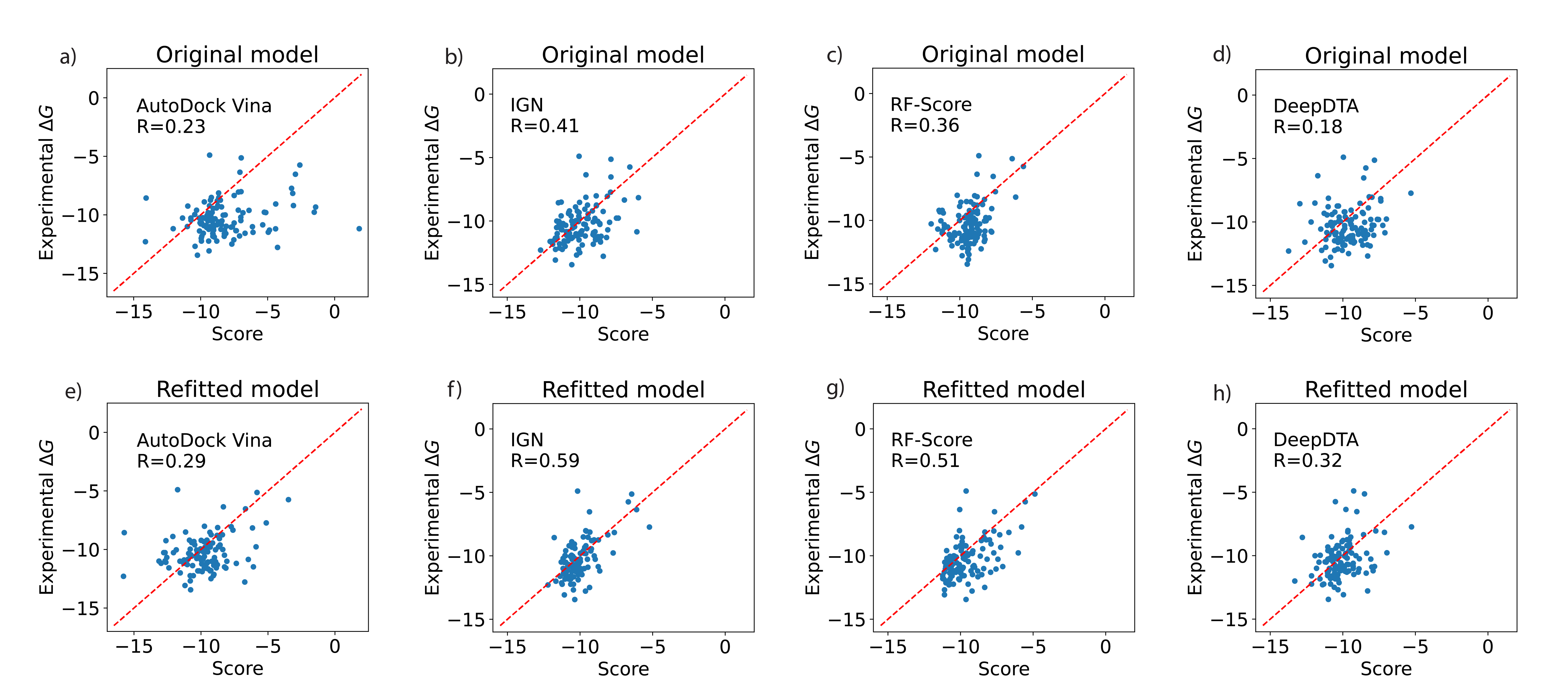}
\end{center}
\caption{\textit{Scatter plots for predicted scores and experimental binding affinities before and after model retraining for different models evaluated on BDB2020+ benchmark dataset.} Scatter plots for original models: AutoDock vina (a), IGN (b), RF-Score (c) and DeepDTA(d); Scatter plots for models retrained with UCBSplit: AutoDock vina (e), IGN (f), RF-Score (g) and DeepDTA (h)}
\label{fig:bdb_scatters}
\end{figure}

\begin{figure}[H]
\begin{center}
\includegraphics[width=1.05\textwidth]{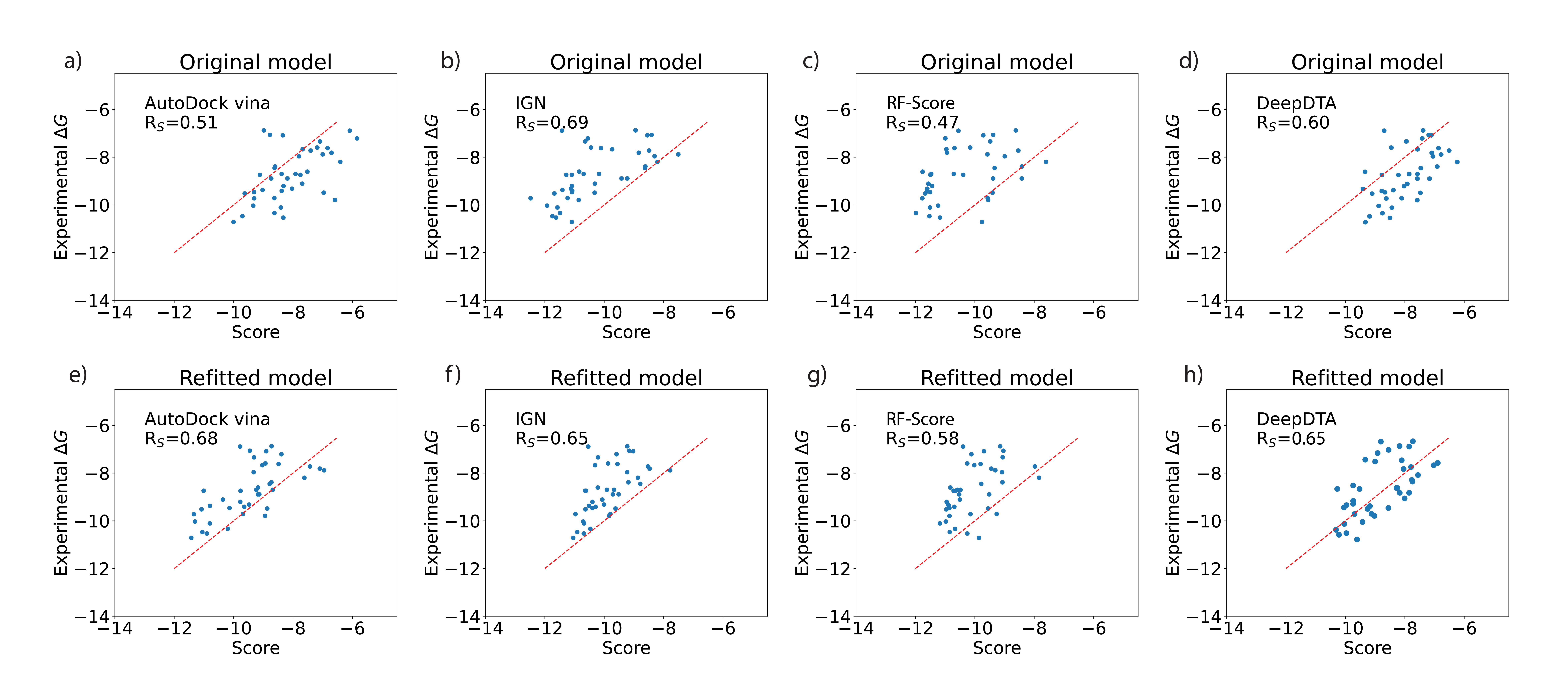}
\end{center}
\caption{\textit{Scatter plots for predicted scores and experimental binding affinities before and after model retraining for different models evaluated on SARS-CoV-2 main protease (Mpro) benchmark dataset.} Scatter plots for original models: AutoDock vina (a), IGN (b), RF-Score (c) and DeepDTA(d); Scatter plots for models retrained with UCBSplit: AutoDock vina (e), IGN (f), RF-Score (g) and DeepDTA (h)}
\label{fig:mpro_scatters}
\end{figure}

\begin{figure}[H]
\begin{center}
\includegraphics[width=1.05\textwidth]{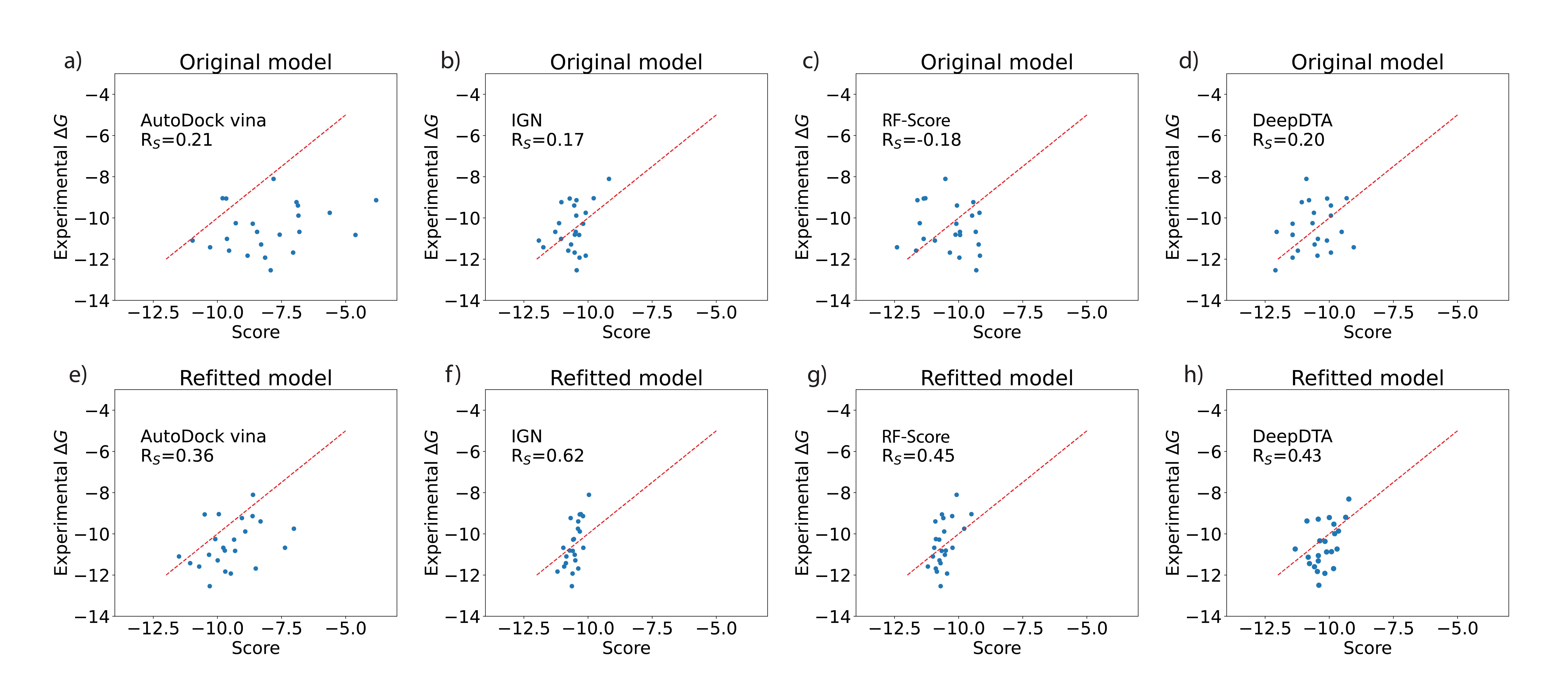}
\end{center}
\caption{\textit{Scatter plots for predicted scores and experimental binding affinities before and after model retraining for different models evaluated on epidermal growth factor receptor (EGFR) benchmark dataset.} Scatter plots for original models: AutoDock vina (a), IGN (b), RF-Score (c) and DeepDTA(d); Scatter plots for models retrained with UCBSplit: AutoDock vina (e), IGN (f), RF-Score (g) and DeepDTA (h)}
\label{fig:egfr_scatters}
\end{figure}
\bibliography{references}